\documentclass[a4paper,fleqn,usenatbib]{mnras}
\usepackage{natbib}

\usepackage{newtxtext,newtxmath}

\usepackage[T1]{fontenc}
\usepackage{ae,aecompl}

\usepackage{graphicx}	
\usepackage{amsmath}	
\usepackage{amssymb}	
\usepackage{booktabs}
\usepackage{color}

\usepackage [english]{babel}
\usepackage [autostyle, english = american]{csquotes}
\MakeOuterQuote{"}
\usepackage[caption=false]{subfig}
\usepackage{caption}

\title[Vortex and flux tube arrays in a neutron star]{Stability of interlinked neutron vortex and proton flux tube arrays in a neutron star: equilibrium configurations}

\author[L. V. Drummond and A. Melatos]{
L. V. Drummond\thanks{E-mail: l.drummond@student.unimelb.edu.au}
and A. Melatos\thanks{E-mail: amelatos@unimelb.edu.au}
\\
School of Physics, University of Melbourne, Parkville, VIC 3010, Australia\\
Centre of Excellence for Gravitational Wave Discovery (OzGrav; CE170100004)\\
}

\date{Accepted XXX. Received YYY; in original form ZZZ}

\pubyear{2016}

\begin{document}

\makeatletter
\let\jnl@style=\rm
\def\ref@jnl#1{{\jnl@style#1}}
\def \rpp {\ref@jnl{Rep.~Prog.~Phys.}}    
\makeatother

\label{firstpage}
\pagerange{\pageref{firstpage}--\pageref{lastpage}}
\maketitle

\begin{abstract}
Three-dimensional, Gross-Pitaevskii equation (GPE) simulations are presented of the interaction between neutron superfluid vortices and proton superconductor flux tubes in a rotating, harmonic trap, representing an idealised model of  the outer core of a neutron star. Low-energy states of the neutron condensate are calculated by evolving the GPE in imaginary time in the presence of a prescribed, static, rectilinear flux tube array. The calculations are carried out as a function of the angle between the global magnetic and rotation axes, and the amplitude and sign of the current-current and density couplings between the neutron and proton condensates. It is found that the system is frustrated by the competition between vortex-vortex repulsion and vortex-flux-tube attraction (pinning), leading to the formation of vortex tangles and "glassy" behaviour characterized by multiple metastable states spaced closely in energy. The dimensionless parameters in the simulations are ordered as one expects in a neutron star, but the dynamic range is many orders of magnitude smaller than in reality, so caution must be exercised when assessing the astrophysical implications. Nevertheless the results suggest that tangled vorticity may be endemic in neutron star outer cores.\end{abstract}

\begin{keywords}
dense matter -- stars: neutron -- stars:rotation  -- stars:interiors -- stars: magnetic field -- pulsars:general 
\end{keywords}



\section{Introduction}


The outer core of a neutron star (density $\rho \sim 10^{14\mbox{--}15} \ \mbox{g }\mbox{cm}^{-3}$) is believed to consist of three interpenetrating fluids: superfluid neutrons, superconducting protons and viscous electrons \citep{Yakovlev1999,LivingReview}. The protons, electrons and rigid crust corotate, while the angular velocity of the neutrons is determined by the number and disposition of the superfluid vortices, each of which carries a quantum of circulation, $\kappa= h/2m_n=1.98 \times 10^{-3} \  \mbox{cm}^2 \ \mbox{s}^{-1}$. The proton superconductor is usually regarded as type II in at least part of the outer core, implying that the magnetic field is concentrated into flux tubes, each carrying a magnetic flux quantum $\Phi_0= hc/2e=2.07 \times 10^{-7}\ \mbox{G} \ \mbox{cm}^{2}$ \citep{,Mendell1991,ChauDing,Field1995}. In this paper we investigate the complex microscopic interaction between neutron vortices and proton flux tubes and relate the results to the star's macroscopic, observable properties, including its angular velocity and magnetic dipole moment. Even though protons constitute $\lesssim 5\%$ of  the outer core by mass \citep{ChauDing,Yakovlev1999}, they play an important role in the stellar rotational dynamics by coupling to the neutrons. For example, neutron vortices may pin to flux tubes, thereby storing angular momentum for later release in rotational glitches  \citep{Sauls1989, Bhatta1991,Ruderman1998,Guger2014}.

A uniformly rotating, axisymmetric superfluid in its minimum energy state is threaded by a rectilinear array of vortices. Similarly, a type-II superconductor immersed in a uniform magnetic field is threaded by a rectilinear array of flux tubes. In the conventional picture of neutron star interiors, the two arrays are assumed to have one or more of the following properties, motivated partly by terrestrial experiments and partly by theoretical convenience: (i) the vortices and flux tubes are straight or gently curved (radius of curvature $\sim$ stellar radius) rather than microscopically tangled; (ii) the vortices and flux tubes are (anti)parallel; and (iii) the interaction energy is large ($\sim$ MeV per vortex-flux-tube junction), so the vortices and flux tubes lock together and move radially outwards in concert, as the neutron star spins down \citep{Bhatta1991}.

In this paper, starting from first principles, we test the above assumptions with the aid of numerical Gross-Pitaevskii equation (GPE) simulations. Specifically, we ask whether the arrays are rectilinear in equilibrium on microscopic and macroscopic scales, when the star's magnetic and spin axes are misaligned. The equilibrium calculations pave the way towards asking in future work, whether the arrays maintain their integrity or are disrupted into interpenetrating tangles under far-from-equilibrium conditions, as the star spins down. These questions are motivated by the observation of tangled vorticity driven by an axial counterflow in hydrodynamic simulations of the outer core \citep{Peralta2005,Peralta2006,AnderssonSidery2007,Howitt2016}, the recognition that the magnetic geometry is complicated (e.g. a closed torus) inside a decelerating star \citep{Easson1979,Melatos2012,GlampLasky}, and the desire to calculate from first principles the driving mechanism underlying the plate tectonic processes proposed by \citet{Ruderman1998}.

There are two general ways to study the above questions: energetics (propagating the GPE in imaginary time to find a low-energy state for a given set of parameters) and dynamics (evolving the GPE in real time, under the action of a driving force which injects energy). We focus on the former approach in this paper and postpone the latter to future work. In a set of controlled numerical experiments, we calculate the equilibrium configurations of interlinked neutron vortex and proton flux tube arrays for realistically ordered values of the dimensionless parameters describing a neutron star. In particular, we preserve the ordering of the strengths of the neutron-neutron self-attraction, neutron-proton density coupling and neutron-proton current-current interaction, while working with less extreme parameter ratios than in a realistic neutron star in order to keep the computations tractable. For the idealised model presented in this paper, we find that an initially rectilinear vortex array bends macroscopically and tangles microscopically in certain regimes. As this happens even for the low-energy equilibria studied here, it can be imagined that driving the system via a spin-down torque is likely to exacerbate the tangling. The latter, far-from-equilibrium dynamics will be studied in a forthcoming article.

The paper is organised as follows. Section 2 outlines how our numerical model treats the neutron fluid, the proton fluid and their interaction. Section 3 elucidates the basic physics of the interaction through the simplified but instructive special case of a neutron vortex array coupled to a single flux tube in two dimensions. The equilibrium configuration of the neutron vortices is calculated as a function of the density and current-current coupling strengths as well as the angle between the macroscopic rotation and magnetic axes. Section 4 extends the results to an array of flux tubes in three dimensions, under conditions where a vortex tangle can form. The energetics, vortex length and mean curvature of the tangle are computed under various conditions, to diagnose the formation of the vortex tangle. In Section 5, we discuss the implications of the results for the macroscopic dynamics and observable properties of the star and point the way to future work, including real-time GPE evolution driven by a spin-down torque as discussed above.

\section{Gross-Pitaevskii Model of the Outer Core}
    \label{section:GPmodel}
\subsection{Neutron superfluid}

We model the neutron superfluid as a dilute Bose-Einstein condensate described by the GPE in line with previous work \citep{LilaMelatos2011, LilaMelatos2012,Douglass2015}. The model has been successfully applied to the superfluid neutron star interior to describe aspects of rotational glitches \citep{LilaMelatos2011}. Nevertheless, it is idealised for the purposes of numerical tractability (see Section \ref{ref:limitations} for details). We emphasize that it neglects many vital aspects of neutron star physics, among them the following: (i) The neutrons are not really dilute, in the sense that the s-wave scattering length is comparable to the average neutron separation \citep{Yakovlev1999,LivingReview}. (ii) The viscous electron plasma in the outer core and the hydromagnetic forces acting upon it (and the protons) are omitted \citep{Mendell1998, Glamp2011}. (iii) We neglect pinning in the inner crust, which varies radially with density and temperature over a length-scale comparable to the inner crust's thickness \citep{Donati2006,Avogadro2007,Avogadro2008,Grill2012,Haskell2013}.

We follow the practice of coupling the condensate to a thermal reservoir \citep{LilaMelatos2011, LilaMelatos2012,Douglass2015}. Thermal coupling promotes numerical stability and simultaneously serves as a proxy for the enhanced neutron scattering occurring in a nondilute condensate. To include it, we solve the semi-quantitative "stochastic GPE" described by \citet{GardinerStochastic} and implemented by \cite{Douglass2015}, which generalises the phenomenological method proposed by \citet{KasamatsuUeda}; see also \citet{LilaMelatos2011} and \citet{LilaMelatos2011}. The stochastic GPE takes the dimensionless form

\begin{equation} 
\label{eq:realGP} 
(\mathrm{i}-\gamma)\frac{\partial\psi}{\partial t}	= \left(-\frac{1}{2}\nabla^{2}+V+|\psi|^{2}-\Omega \hat{L}_{z}+\mathrm{i}\gamma \mu \right)\psi+\mathcal{H}_{int}[\psi,\phi],
\end{equation} 
where $\psi$ is the neutron condensate order parameter, and $\phi$ is the proton condensate order parameter. We normalise $\psi$ to give $\int|\psi|^2\mbox{d}^3\mathbf{x}=N_n$ and $\phi$ to give $\int|\phi|^2\mbox{d}^3\mathbf{x}=N_p=0.05N_n$ \citep{LivingReview}, where $N_n$ and $N_p$ are the total numbers of condensed neutrons and protons respectively. Appendix \ref{section:modification} discusses possible modifications of the kinetic energy due to entrainment.

The term $\mathrm{i}\gamma\mu\psi$ in (\ref{eq:realGP}) suppresses sound waves, where $\gamma$ is a damping constant (the value of which is chosen based on semi-quantitative analogies with terrestrial experiments) and $\mu$ is the chemical potential of the uncondensed fraction. We obtain $\mu$ through imaginary time evolution, because one has $\psi \propto e^{-\mathrm{i}\mu t/\hbar}$ in the rotating frame, i.e. $\mu$ is the chemical potential of the equilibrium state. Further details concerning the physical meaning of $\mu$ and the value of $\gamma$ in a neutron star superfluid are presented elsewhere \citep{LilaMelatos2011, Douglass2015}.

The dimensionless independent variables in equation (\ref{eq:realGP}) are normalized relative to characteristic length- and time-scales $\xi_n=\hbar/(2m_nn_{n}U_{0})^{1/2}$ and $\tau=\xi_n/c_{s}=\hbar/(n_{n}U_{0})$ respectively, where $U_{0} $ is the strength of the neutron self-interaction, $n_n$ is the background neutron number density, $\xi_n$ is the neutron coherence length and $c_s$ is the speed of sound. The dependent variables $\psi$ and $\phi$ are both expressed in units of $n_n^{1/2}$. Equation (\ref{eq:realGP}) is written in the frame rotating with the same angular velocity $\Omega$ as the crust, and $\hat{L}_{z}$ is the angular momentum operator. The neutrons are trapped in a harmonic potential $V$ with cylindrical geometry, viz. $V=\tilde{\omega}^{2}(x^{2}+y^{2})/2$ in Cartesian coordinates, where $\tilde{\omega}$ is the trap frequency normalised by $\tau^{-1}$.\footnote{The form of the harmonic trap is broadly consistent with the potential resulting from hydrostatic equilibrium in a neutron star ($V\propto r^{2}$ for a self-gravitating star with constant density).} Interaction terms are included via $\mathcal{H}_{int}[\psi,\phi]$, a functional of $\psi$ and $\phi$, whose physical and mathematical form we discuss in Section 2.3. The neutron superfluid is coupled to the flux tubes via both density and current-current interactions. 

In this article, we calculate the low-energy states of (\ref{eq:realGP}) by solving the imaginary-time GPE ($t\,\to\,-\mathrm{i}t$, $\gamma\to0$)

\begin{equation} \label{eq:imagGP} \frac{\partial\psi}{\partial t}	= \left(\frac{1}{2}\nabla^{2}-V-|\psi|^{2}+\Omega \hat{L}_{z}\right)\psi-\mathcal{H}_{int}[\psi,\phi]
\end{equation} 
to obtain the equilibrium wavefunction $\psi_{gs}$ as $t\,\to\,\infty$. Real-time solutions of equation (\ref{eq:realGP}) with a driving force will be discussed in a future article. Strictly speaking, the imaginary-time procedure yields the ground state, but convergence to the minimum energy is only guaranteed after an infinitely long time. The imaginary-time evolution of a frustrated ("glassy") system with competing interactions is protracted, as the system vacillates between intermediate low-energy states with almost the same energy, before the true ground state is reached \citep{Anderson2004,sibani2013}. The systems we investigate in this article are generally glassy and exhibit this behaviour, as we show in detail in Section \ref{section:FTA}.

\subsection{Proton superconductor}
\label{section:protonSC}
We model the superconducting protons using Ginzburg-Landau theory with the standard minimal electromagnetic coupling prescription $\nabla	\rightarrow	\nabla-2\mathrm{i}e\mathbf{A}/\hbar c$. We assume that $\phi$,  the proton order parameter, and $\mathbf{A}$, the magnetic vector potential, are smooth \citep{Tinkham} and obey the coupled, Ginzburg-Landau equations given by \citep{Tinkham} 

\begin{equation}
 \label{eq:GL1} 
\mathrm{i}\hbar\frac{\partial\phi}{\partial t}	= \frac{1}{4m_p}\left(-\mathrm{i}\hbar\nabla-\frac{2e\mathbf{A}}{c}\right)^{2}\phi+\alpha\phi+\beta|\phi|^2\phi,
\end{equation}

\begin{equation}
 \label{eq:GL2} 
\nabla\times\mathbf{B}= \frac{4\pi e \mathbf{j}_p}{m_pc},
\end{equation}
with
\begin{equation}
 \label{eq:GL3} 
\mathbf{j}_p=\frac{\mathrm{i\hbar}}{2}\left[\phi\left(\nabla+\mathrm{i}\frac{2e}{\hbar c}\mathbf{A}\right)\phi^{*}-\phi^{*}\left(\nabla-\mathrm{i}\frac{2e}{\hbar c}\mathbf{A}\right)\phi \right].
\end{equation}
In (\ref{eq:GL1})--(\ref{eq:GL3}), $\alpha$ and $\beta$ are phenomenological parameters \citep{Gorkov}, $m_p$ is the proton mass and $\mathbf{j}_p$ is the proton momentum density. Equations (\ref{eq:GL1})--(\ref{eq:GL3}) are written in dimensional form.

In this paper, we adopt an ansatz for a static flux tube array instead of solving equations (\ref{eq:GL1})--(\ref{eq:GL3}) for $\phi$ and $\mathbf{A}$ directly. This approximate solution has the same qualitative behaviour as numerical solutions of the Ginzburg-Landau equations \citep{Clem}. Unlike the London model, where $\mathbf{A}$ diverges on the vortex axis \citep{Clem,Mendell1991}, this ansatz gives realistic values inside the vortex core. For an isolated flux tube at the origin in polar co-ordinates $(r,\chi)$ we have \citep{Clem}

 \begin{equation}
  \label{eq:phi} 
 \phi_f=n_p^{1/2}(r/\tilde{r})^2\mbox{e}^{\mathrm{i}\chi},
  \end{equation}
  
   \begin{equation}
     \label{eq:Af} 
 \mathbf{A}_f 	=	\frac{\Phi_{0}}{2\pi r} \left[1-\frac{\tilde{r}K_{1}\left(\tilde{r}/\lambda\right)}{\sqrt{2}\xi_p K_{1}\left(\sqrt{2}\xi_p/\lambda \right)} \right] \hat{\chi},
  \end{equation}
 in dimensional form, where $\xi_p$ is the proton coherence length, $\lambda$ is the London penetration depth, we write $\tilde{r}=(r^2+2\xi_p^2)^{1/2}$, and $K_{n}(x)$ is a modified Bessel function of the second kind of order $n$. Expressions for $\xi_p$ and $\lambda$ that account for entrainment are given in Appendix \ref{section:londoncoherence}.
 
The flux tube array is constructed from (\ref{eq:phi}) and (\ref{eq:Af}) by forming the product of single-flux-tube wavefunctions $\phi(\mathbf{x})=\prod_{i}\phi_f(\mathbf{x}-\mathbf{x}_i)$ and a linear superposition of the single-flux-tube vector potentials $\mathbf{A}(\mathbf{x})=\sum_{i}\mathbf{A_f}(\mathbf{x}-\mathbf{x}_i)$, where $\mathbf{x}_i$ is the position of the $i^{\mbox{th}}$ flux tube in the mid-plane of the system. The above product and sum are good approximations, when the flux tubes are sufficiently well separated. The internal density profile of each flux tube can be considered approximately independent, provided that the distance between flux-tube cores exceed $\approx 5\xi_p$ \citep{BrandtRev}.\footnote{Each flux tube is described by $\phi_f$ and $\textbf{A}_f$ profiles and a location $\mathbf{x}_i$. The profiles are independent, when the flux tubes are well-spaced, but the locations are correlated due to long-range, mutually repulsive interactions between flux tubes which lead to the formation of a triangular Abrikosov lattice.} This is true in a neutron star, where vortices are spaced by $d_{v}=3.4\times10^{-3} \ (\Omega/10^2\ \mbox{rad} \ \mbox{s}^{-1})^{-1/2} \ \mbox{cm}$, while flux tubes are spaced by $d_{\Phi}=3\times10^{-10}\ (B/10^{12} \ \mbox{G})^{-1/2}\ \mbox{cm}$ \citep{Link2003}. The neutron superfluid mimics rigid rotation by establishing a lattice of vortices with area density $n_{v}= 2\Omega/\kappa=d^{-2}_{v}$ directed along $\boldsymbol{\Omega}$. The proton superconductor contains an array of flux tubes with area density $n_{\Phi}=B/\Phi_{0}=d^{-2}_{\Phi}$ directed along $\mathbf{B}$. In a realistic neutron star, one has $n_{\Phi}/n_{v}\sim10^{13}$ \citep{LivingReview} As we cannot capture such a wide dynamic range numerically, we set $2\leq n_{\Phi}/n_v\leq 5$ in our calculations.

 Dynamo processes during the formation of the neutron star arguably bestow a complicated large-scale structure on the magnetic field, which is retained as the proton fluid condenses into a superconductor and is maintained by the proton-neutron lag as the star spins down \citep{Easson1979,Ruderman1998,JahanMiri, Melatos2012,GlampLasky}. Although type-I superconductivity is less likely than type-II by analogy with a two-gap superconductor, nonetheless the protons may be type-I in some regions of the outer core, where $B$ is high enough to quench the superconductivity \citep{RudermanUnsolved,Jones2006}.

\begin{figure}
	\includegraphics[width=\columnwidth]{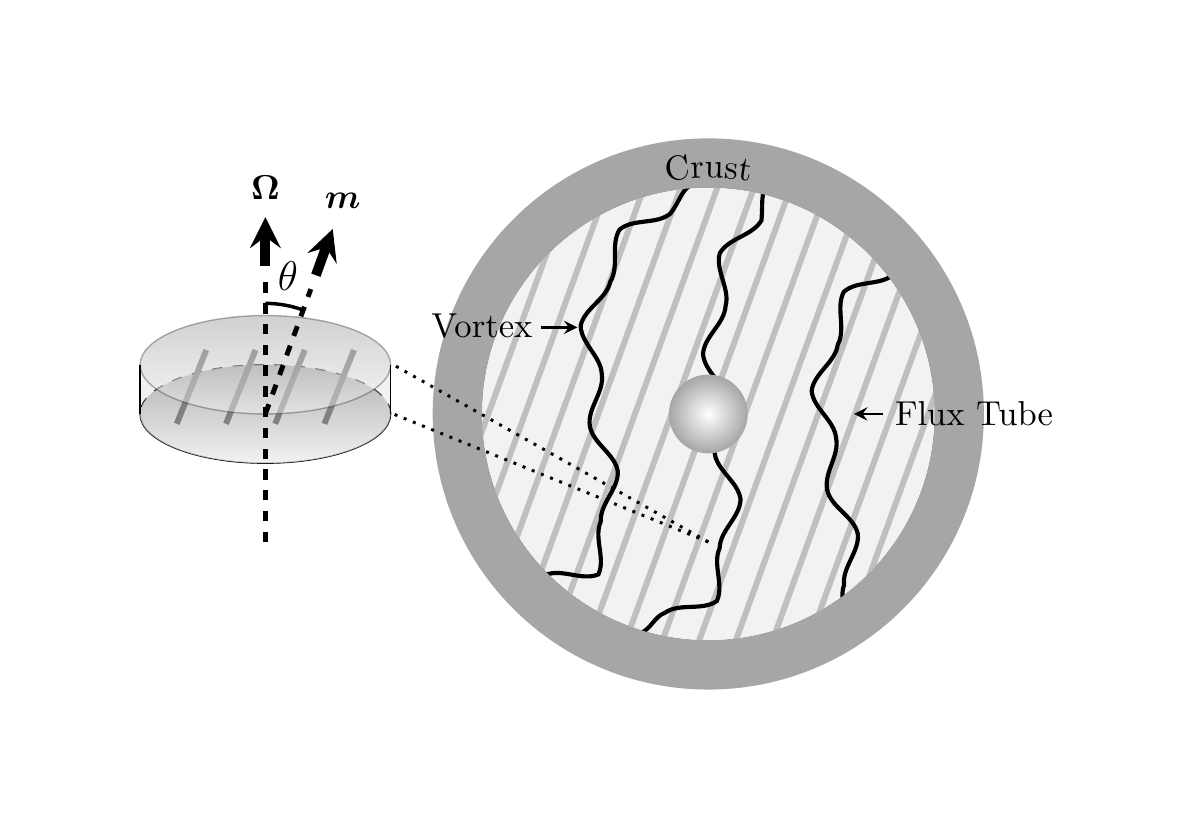}
    \caption{Schematic of a possible, realistic configuration of the superfluid neutron vortices and superconducting proton flux tubes in the neutron star outer core. The vortices lie broadly parallel to the rotation axis $\boldsymbol{\Omega}$. They form a tangled, frustrated equilibrium as they seek a compromise between aligning with $\boldsymbol{\Omega}$ globally and pinning to the flux tubes locally. The flux tubes align roughly with the global magnetic dipole moment $\mathbf{m}$ but also have a complicated geometry locally \citep{Easson1979,Melatos2012,GlampLasky}, which we do not model in this paper. We solve the GPE in a "box" of fluid centred on the rotation axis (inset). The box is small enough that the flux tubes can be approximated as straight inside it and inclined at an angle $\theta$ with respect to $\boldsymbol{\Omega}$. The rigid crust (dark shaded annulus) is not modelled explicitly in this paper; nor is the inner core (shaded circle at centre).}
    \label{fig:stardiagram}
\end{figure}

A schematic of the possible structure of the star is depicted in Figure \ref{fig:stardiagram}. In this paper, we assume the flux tubes are rectilinear and static, while the vortices are free to adopt a complicated geometry as they struggle to equilibrate in the presence of competing forces, aligning with the rotation axis $\boldsymbol{\Omega}$ globally and pinning to the flux tubes locally. The flux tube response will be studied in a future paper. In this paper, we simulate a microscopic box centred on the rotation axis, as indicated in the inset of Figure \ref{fig:stardiagram}. The flux tubes in the box are inclined at an arbitrary angle $\theta$ with respect to $\boldsymbol{\Omega}$. An on-axis box is chosen to capture qualitatively the effect of the rotation and axial geometry. In reality one needs a macroscopically sized box to capture rotational effects faithfully (e.g. outward vortex drift during spin-down), but a macroscopic box is out of reach computationally at present. 

\subsection{Interaction between the components}
\label{subsection:interaction}
In this work, we investigate how the neutrons respond dynamically to the steady-state, prescribed by proton ansatz in Section 2.2. The fully coupled system, where neutrons and protons are both free to evolve dynamically, will be examined in a future paper. 

The neutrons and protons interact through $\mathcal{H}_{int}$ in equation (\ref{eq:realGP}). The interaction manifests itself in several ways both in the bulk and at the sites of topological defects. One of the most important manifestations is pinning, where topological defects overlap and "stick" to each other at locations they would not occupy with $\mathcal{H}_{int}=0$. It is energetically favourable for the flux tubes and vortices to pin for several reasons \citep{Bhatta1991, Ruderman1998}. We study two pinning mechanisms here. (i) \textit{Density interaction:} it is energetically favourable for the density minima in the cores of a vortex and a flux tube to overlap. (ii) \textit{Current interaction:} the neutron and proton momentum densities interact to produce an entrainment effect. Mechanism (ii) causes neutrons to drag protons along as they circulate, generating a magnetic field with strength $B_n\sim10^{14}\,\mbox{G}$ within a vortex core, illustrated in Figure \ref{fig:crossingenergy}, which interacts electromagnetically with the magnetic field $B_p\sim10^{15}\,\mbox{G}$ in a flux tube, adding to the strength of the current-current interaction. Mechanisms (i) and (ii) are described by phenomenological terms in the GPE Hamiltonian \citep{AlparSaulsLanger, AnthonyThesis},

\begin{equation}
\label{eq:Hint} \mathcal{H}_{int}[\psi,\phi]=\eta|\phi|^{2}\psi - \frac{\mathrm{i}\zeta}{2}\left(2\mathbf{j}_p\cdot\nabla\psi+\psi\nabla\cdot\mathbf{j}_p\right), 
\end{equation}
with
 \begin{equation}
 \label{eq:nodimjp}
\mathbf{j}_p=\frac{\mathrm{i}}{2}\left[\phi\left(\nabla+\mathrm{i}\xi_{n}\frac{2e}{\hbar c}\mathbf{A}\right)\phi^{*}-\phi^{*}\left(\nabla-\mathrm{i}\xi_{n}\frac{2e}{\hbar c}\mathbf{A}\right)\phi \right]
 \end{equation}
where $\eta$ and $\zeta$ are the dimensionless density and current-current coupling coefficients respectively, and (\ref{eq:nodimjp}) is the dimensionless version of (\ref{eq:GL3}). 

The total potential energy due to the density and current-current interactions is \citep{AlparSaulsLanger,AlfordGood2008}

\begin{equation}
 \label{eq:Eint1}
E_\mathrm{int}=\int \mbox{d}^3  \mathbf{x} \ \left( U_0 \eta |\psi|^2|\phi|^2 +\frac{\zeta \mathbf{j}_n \cdot \mathbf{j}_p}{2m_n n_n}\right),
\end{equation}
with
\begin{equation}
\mathbf{j}_n=\frac{\mathrm{i}}{2}\left(\psi\nabla\psi^{*}-\psi^{*}\nabla\psi \right),
\end{equation}
where $E_\mathrm{int}$ is dimensional and expressed in terms of dimensionless quantities $\eta$, $\zeta$, $\mathbf{j}_n$ and $\mathbf{j}_p$.
We can use (\ref{eq:Eint1}) to relate $\eta$ and $\zeta$ to published formulas for the pinning energy and hence express $\eta$ and $\zeta$ in terms of neutron star parameters. The pinning energy (i.e. the energy difference between the pinned and free configurations) per vortex-flux-tube junction (volume $\xi_n^2\xi_p$) due to the density interaction is given by $E_{\eta}= n_n \xi_{n}^{2} \xi_{p} \Delta_{p}^{2}\Delta_{n}^{2}/(E_{Fp}^{2}E_{Fn})$, where the symbols $\Delta$ and $E_{F}$ denote the energy gap and Fermi energy respectively \citep{Sauls1989,Srivasan1990,Bhatta1991,Ruderman1998}. This formula is broadly consistent with the results in \citet{SinhaSedrakian}, which agree with the mean-field calculation in \citet{AlfordGood2005} except for minor typographical errors. Equivalently, from (\ref{eq:Eint1}) we also have $E_{\eta}= U_0 \eta n_n n_p \xi_n^2\xi_p$. Equating the two expressions for $E_{\eta}$ yields
{\begin{align}
 \label{eq:eta}
\eta  &= \left(\frac{n_n}{n_p}\right) \left(\frac{\Delta_{p}}{E_{Fp}}\right)^2\left(\frac{\Delta_{n}}{E_{Fn}}\right)\frac{2m_n\xi_n^2\Delta_{n}}{\hbar^2} \\
&=0.2 \left(\frac{\Delta_{p}}{1\ \mbox{MeV}}\right)^2 \left(\frac{\Delta_{n}}{0.1\ \mbox{MeV}}  \right)^2  \nonumber \\
 & \ \ \ \times\left(\frac{E_{Fp}}{3\ \mbox{MeV}}\right)^{-2} \left(\frac{E_{Fn}}{60\ \mbox{MeV}}\right)^{-1}   \left(\frac{\xi_n}{100\ \mbox{fm}}\right)^2. 
\end{align}
In the neutron star outer core one typically has $\Delta_{n}=0.1\ \mbox{MeV}$ and $\Delta_{p}=1\ \mbox{MeV}$; the reader is referred to \citet{bandgap}, \citet{Yakovlev1999} and \citet{Beloin2016} for details. One also has $E_{Fn}=60\mbox{--}100\ \mbox{MeV}$ and $E_{Fp}=3\mbox{--}6\ \mbox{MeV}$ \citep{Shapiro,Yakovlev1999}.  Bardeen-Cooper-Schrieffer calculations yield consistent values for $\xi_n$; see Eq. (17)  in \citet{AlparSaulsLanger} and Eq. (A8) in \citet{Mendell1998}. We note that $\eta$ is believed to be negative (i.e. an attractive "pinning" interaction) according to first principles calculations by \citet{AlfordGood2005}.

\begin{figure}
\centerline{	\includegraphics[width=0.8\columnwidth]{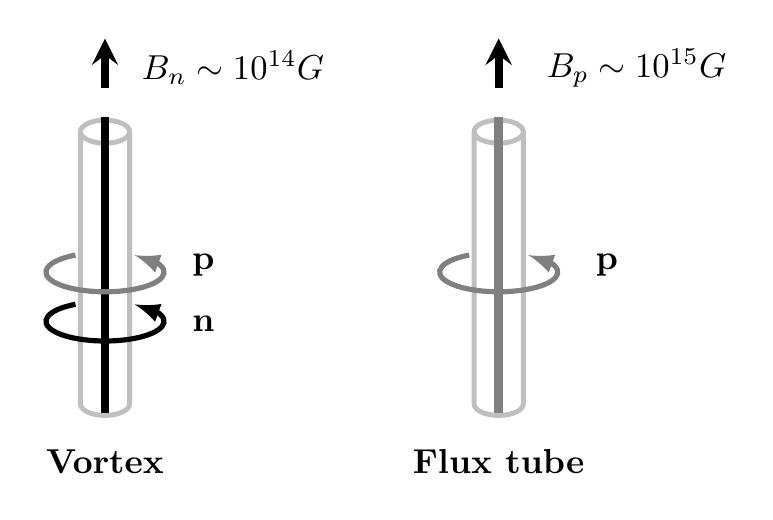} }
    \caption{The entrainment of protons by circulating neutrons endows a neutron vortex with a magnetic field $B_n$. Neutrons are also entrained by protons circulating around a flux tube but, being neutral, they do not contribute to its magnetic field $B_p$.}
    \label{fig:crossingenergy}
\end{figure}

\begin{figure}
	\centerline{ \includegraphics[width=0.8\columnwidth]{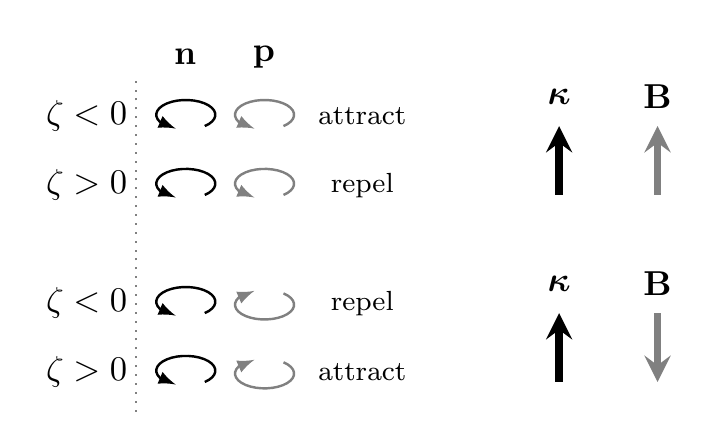} }
    \caption{The sign and magnitude of the current-current coupling depends on the orientation of the local magnetic field $\mathbf{B}$ relative to the local circulation $\mathbf{\kappa}$.}
    \label{fig:entrainment}
\end{figure}

The current-current coupling ("entrainment") parameterised by the dimensionless coefficient $\zeta$ arises fundamentally from the strong nuclear force between the neutrons and protons. We estimate the size of $\zeta$ from entrainment parameters published elsewhere in the literature. According to the hydrodynamical description given in \citet{AlparSaulsLanger}, based on pioneering work by \citet{Andreev1975}, entrainment contributes a term of the form $\rho_{pn} \mathbf{v}_n \cdot \mathbf{v}_p$ to the potential energy density, where $\mathbf{v}_n$ and $\mathbf{v}_p$ are the neutron and proton bulk velocities respectively and $\rho_{pn}$ is an entrainment coefficient. Equating the latter quantity with the energy density $\zeta \mathbf{j}_n \mathbf{j}_p/(2m_n n_n)$ in (\ref{eq:Eint1}), we obtain $\zeta=\rho_{pn}/(2m_nn_p)$ \footnote{We observe that $\rho_p=2m_p n_p$, because $n_p$ is the number density of proton Cooper pairs.}. Calculations involving Bardeen-Cooper-Schrieffer theory and Fermi-liquid theory \citep{AlparSaulsLanger} yield
\begin{equation}
\label{eq:zeta}
\zeta=\frac{\delta m_p^*}{m_p^*}
\end{equation}
where $m_p^*=m_p+\delta m_p^*$ is the "dressed proton mass" and $\delta m^*$ is the change in the effective proton mass due to the dragged polarisation cloud of neutrons and protons. Estimates based on generalising Landau's effective mass model to a two-component Fermi system give $\delta m_p^*=-0.5m_p$ and hence $\zeta=-0.5$ \citep{Sjoberg1976}. Modern calculations involve density functional theory based on a self-consistent mean-field model \citep{entrainparam,AlfordGood2008}. This approach accounts for the density dependence of entrainment and gives $-1.2\leq\zeta\leq-0.2$ in the outer core [see Figure 2 in \citet{entrainparam}].

In the GPE simulations in this paper, the protons are not free to be entrained by the neutrons. Hence, they do not circulate around vortices to generate an additional magnetic field $B_n$ as in Figure \ref{fig:crossingenergy}. This is because we do not solve for $\phi$ and $\mathbf{A}$ but adopt a static ansatz instead. The entrained proton mass current we omit in this work is of the form $(\zeta/2m_nn_p)m_p|\phi|^2(\mathrm{i}\hbar/2)\left(\psi\nabla\psi^{*}-\psi^*\nabla\psi\right)$, which generates a magnetic field via (\ref{eq:GL2}). This magnetic field contributes to the energy of the system by an amount $E_{\mathrm{mag}}= g(\theta)B_{n}B_{p}\pi\lambda_{n}^{2}\lambda_{p}/(8\pi)$ at each vortex-flux-tube junction \citep{Bhatta1991, Jones1991, Mendell1991, ChauDing,Ruderman1998,Link2012}, where $g(\theta)$ is a dimensionless function of $\theta$, the angle between the rotation and magnetic axes. The magnetic interaction enhances or reduces the pinning potential depending on the relative orientation of $\boldsymbol{\Omega}$ and $\mathbf{B}$.

The geometric dependence $g(\theta)$ contains a factor $\propto\mbox{cos} \ \theta$ from the dot product of the vortex and flux tube magnetic fields and a factor $\propto (\mbox{sin} \ \theta)^{-1}$ due to the variation in overlap length \citep{Jones1991,ChauDing,Ruderman1998,Link2012}. Although $g(\theta)$ is usually quoted in expressions for $E_{\mathrm{mag}}$ in the literature, a similar factor is expected to enter $\zeta$ for the current-current interaction in general, whether or not $E_{\mathrm{mag}}$ dominates, and we include it henceforth. The effect of $\theta$ on the interaction strength is illustrated in Figure \ref{fig:entrainment}. When the magnetic moment $\mathbf{m}$ and rotation axis $\boldsymbol{\Omega}$ are aligned, the direction of the circulation around vortices and flux tubes is the same. A negative (positive) current coupling $\zeta$ favours (anti) alignment of currents and hence attraction (repulsion) of vortices and flux tubes, as in the top half of Figure \ref{fig:entrainment}. When $\mathbf{m}$ and $\boldsymbol{\Omega}$ are antialigned, the opposite interaction occurs, as in the bottom half of Figure \ref{fig:entrainment}. We verify this numerically in Section \ref{section:SFT}.

In Section \ref{section:SFT} we explore a wide range of $|\zeta|$ and $|\eta|$ values to test how the system's behaviour depends on coupling strength. In Section \ref{section:FTA} we select physically interesting values of $\zeta$ and $\eta$ that are broadly consistent with estimates in this section.

\section{Single flux tube}
\label{section:SFT}
We begin by exploring the interaction between a single flux tube and a vortex array in two dimensions. The situation is not directly relevant to a neutron star, but it lends valuable insight into how the couplings discussed in section 2.3 affect the vortex motion locally. As a control experiment, to set a baseline for what follows, we calculate the ground-state structure of 20 neutron vortices and a single, off-axis, proton flux tube with zero coupling. The density, phase and momentum density are plotted for both fluids in Fig. \ref{fig:nocouplingandfluxtube}. We see from panels (c) and (f) that both fluids circulate clockwise, i.e. the magnetic and rotation axes are aligned and pointing into the page. We confirm that the phase rotates by $2\pi$ around each topological defect in panels (b) and (e).

\begin{figure*}
	\includegraphics[width=\textwidth]{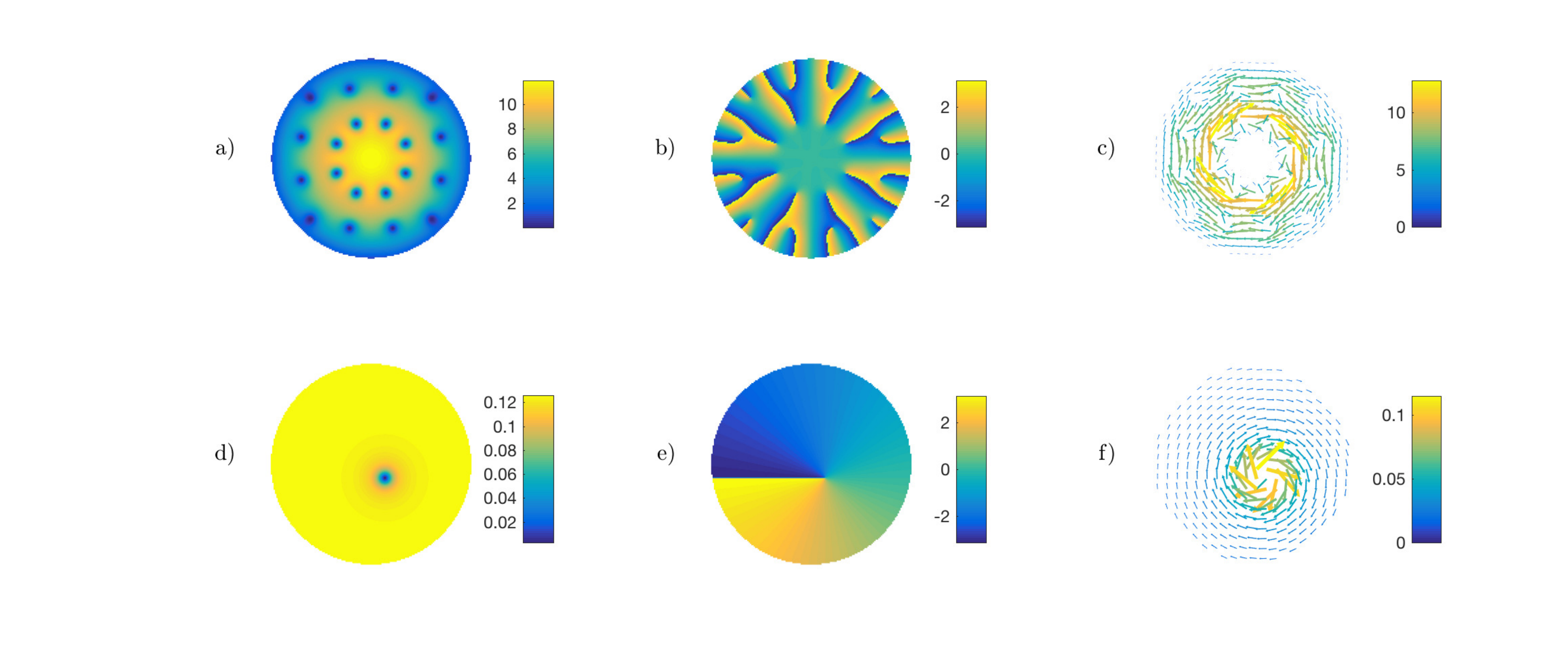}
    \caption{Ground-state structure of a neutron superfluid containing 20 vortices and a proton superconductor containing a single, off-axis flux tube with zero coupling, i.e. $\eta=0$, $\zeta=0$. (a) Neutron density $|\psi|^2$. (b) Neutron phase $\mbox{arg}(\psi)$. (c) Neutron current $\mathbf{j}_n=(\mathrm{i}/2)\left(\psi\nabla\psi^{*}-\psi^{*}\nabla\psi\right)$. (d) Proton density $|\phi|^2$. (e) Proton phase $\mbox{arg}(\phi)$. (f) Proton current $\mathbf{j}_p=(\mathrm{i}/2)\left\{\phi\left[\nabla+\mathrm{i}\xi_{n}(2e/\hbar)\mathbf{A}\right]\phi^{*}-\phi^{*}\left[\nabla-\mathrm{i}\xi_{n}(2e/\hbar)\mathbf{A}\right]\phi\right\}$. A colour bar defines the contours in each figure. The units are $n_n$ for both panels (a) and (d) and $n_n\hbar/\xi_n$ for panels (c) and (f).}
    \label{fig:nocouplingandfluxtube}
\end{figure*}

\subsection{Attractive versus repulsive coupling}
\label{subsection:attractvsrepulse}
 An attractive density coupling $\eta<0$ makes it energetically favourable for density "holes" (i.e. the cores of the topological defects) to overlap. A repulsive density coupling $\eta>0$ makes it favorable for a peak in the neutron density to overlap with a flux tube.  Figure \ref{fig:currentalignment} illustrates how the sign and type of interaction affects the ground-state structure of the neutron vortices. The figure contains three subpanels in each column: the density of the neutron fluid in the presence of a flux tube (top subpanel), where the position of the flux tube is denoted by a red circle; a colour map of the normalised dot product of the proton and neutron currents, $\mathbf{j}_n \cdot \mathbf{j}_p/j_nj_p$ (middle subpanel); and a histogram of $\mathbf{j}_n \cdot \mathbf{j}_p/j_nj_p$ (bottom subpanel), computed by evaluating the dot product at each grid cell (all of which have the same area), excluding points outside the edge of the condensate, and dividing the counts into eight bins. The asymmetry introduced by the offset flux tube accentuates the effect of the coupling. 
 
 An attractive current coupling ($\zeta<0$) tends to align $\mathbf{j}_n$ and $\mathbf{j}_p$, leading to the dragging or entrainment effect discussed in Section \ref{subsection:interaction}. The currents tend to counter-align for $\zeta>0$. In order to demonstrate this, we plot a colour map of $\mathbf{j}_n \cdot \mathbf{j}_p/j_nj_p$ in the middle subpanel of Figure \ref{fig:currentalignment} to visually demonstrate the extent of the alignment, where the yellow (blue) shading denotes alignment (counteralignment). It is important to appreciate that an attractive current coupling favours maximising $\mathbf{j}_n\cdot \mathbf{j}_p$, not $\mathbf{j}_n\cdot \mathbf{j}_p/j_nj_p$, i.e. the system responds to the magnitudes of the currents as well as their directions. 
 
 Looking at the middle subpanel we see less blue colouration in Figure \ref{fig:currentalignment}(a) ($\zeta<0$) compared to Figure \ref{fig:currentalignment}(b) ($\zeta>0$). Similarly, more negative values and a slight peak at $-1$ occur in the $\mathbf{j}_n \cdot \mathbf{j}_p/j_nj_p$ histogram in Figure \ref{fig:currentalignment}(b) compared to Figure \ref{fig:currentalignment}(a). For $\eta<0$, the vortices tend to overlap with the flux tube, with one vortex sitting on the red circle in the top subpanel of Figure \ref{fig:currentalignment}(c). For $\eta>0$, the vortices move away from the red circle, and the neutron density in the region near the flux tube tends to be $\sim60 \%$ higher than the surrounding fluid, as seen in the top subpanel of Figure \ref{fig:currentalignment}(d). The behaviour of the density coupling does not depend on orientation. The $\mathbf{j}_n \cdot \mathbf{j}_p/j_nj_p$ histograms in Figures \ref{fig:currentalignment}(c) and \ref{fig:currentalignment}(d) resemble each other more closely than those in Figures \ref{fig:currentalignment}(a) and \ref{fig:currentalignment}(b) for the current coupling. Nonetheless, antialigned currents are suppressed slightly for $\eta<0$ [see Figure \ref{fig:currentalignment}(c)] because vortex alignment, regardless of what kind of interaction triggers it, causes some current alignment automatically, when the senses of circulation of the neutrons and protons are the same.
 
In some of the equilibria in Figure \ref{fig:currentalignment}, the coupling pushes  some vortices over the edge of the condensate  (drawn where $|\psi|^2$ drops below $10\%$ of its maximum). For example,  Figures \ref{fig:currentalignment}(a)--\ref{fig:currentalignment}(d) contain 19, 16, 19, and 16 vortices respectively, compared to 20 in the $\eta=0$, $\zeta=0$ control experiment in Figure \ref{fig:nocouplingandfluxtube}.
 
\begin{figure*}
	\includegraphics[width=\textwidth]{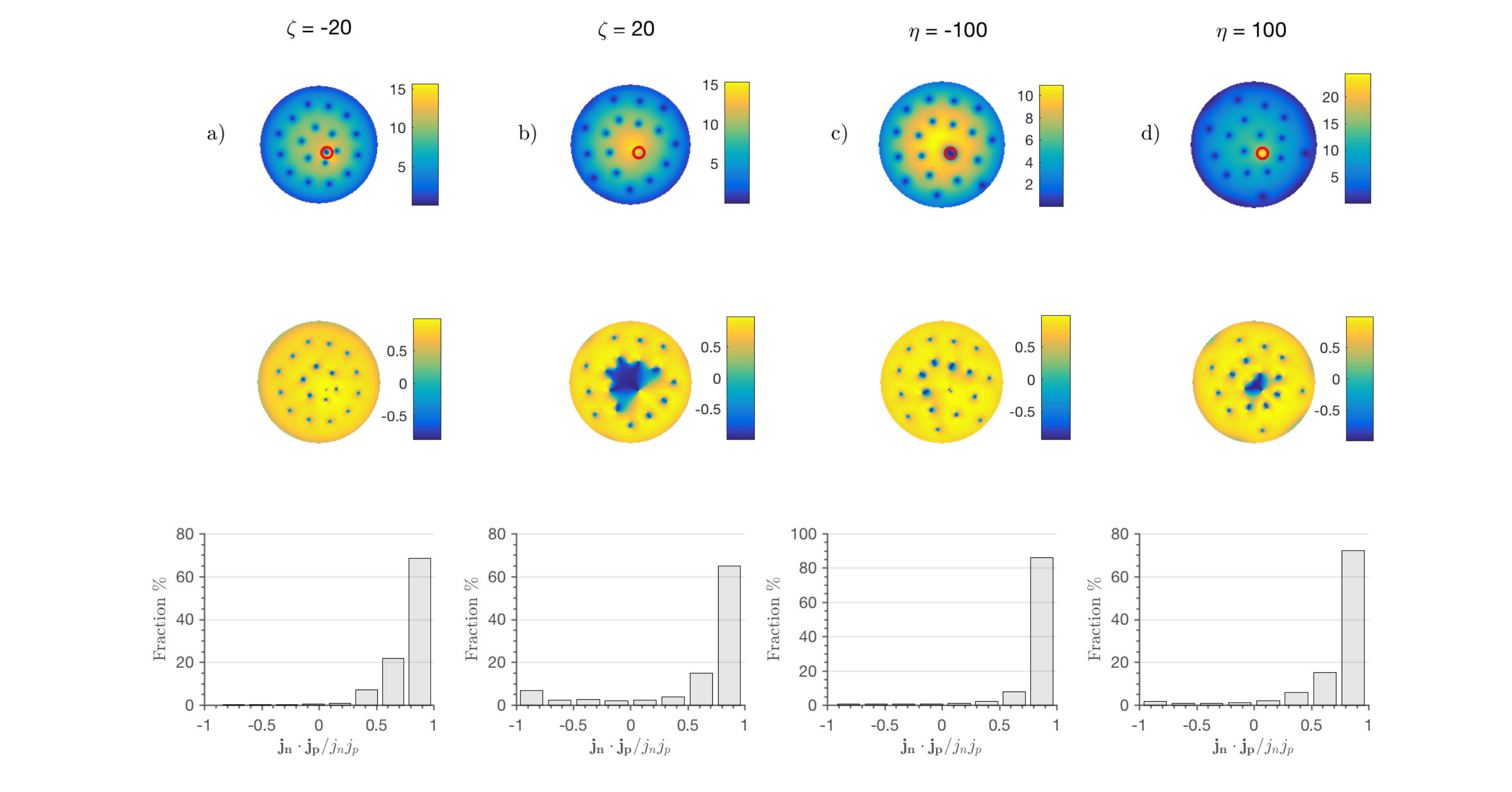}
    \caption{Ground-state structure of a neutron superfluid coupled to a proton superconductor containing a single off-axis flux tube with $\mathbf{B}$ and $\boldsymbol{\Omega}$ aligned. (a) Attractive current coupling $\zeta=-20$. (b) Repulsive current coupling $\zeta=20$. (c) Attractive density coupling $\eta=-100$. (d) Repulsive density coupling $\eta=100$. The top subpanels display the neutron density $|\psi|^2$, the middle subpanels display the normalised dot product of the current vectors $\mathbf{j}_n \cdot \mathbf{j}_p/j_nj_p$ and the bottom subpanels display a histogram of $\mathbf{j}_n \cdot \mathbf{j}_p/j_nj_p$.  A colour bar is provided for the contour values in each panel. The units are $n_n$ in the top subpanels. The red circle in the top subpanel marks the position of the off-axis flux tube. Parameters: $\Omega=0.5$, $\tilde{\omega}^2=0.6$, $\tilde{N}_n=N_n/(n_n\xi^{3})=10^3$.}
    \label{fig:currentalignment}
\end{figure*}

\subsection{Relative orientation}
\label{subsection:orientation}

The forces between vortices and flux tubes due to the density coupling depend only on the sign and magnitude of $\eta$. For the current coupling the forces depend on the sign and magnitude of $\zeta$ as well as the orientation of the vortices relative to the flux tubes. In three dimensions the latter effect is important as it leads to a rich variety of tangled ground states under certain conditions, to be discussed in Section \ref{section:FTA}. In two dimensions, the effect is simpler: if the flux tubes have a circulation in the opposite sense to the vortices, attraction becomes repulsion and vice versa. 

Figure \ref{fig:fliptest} illustrates the above property. It displays $|\psi|^2$ (top subpanel) and $\mathbf{j}_n\cdot \mathbf{j}_p/j_nj_p$ (bottom subpanel) for  $\zeta<0$, $\zeta>0$, $\eta<0$ and $\eta>0$, when $\boldsymbol{\Omega}$ and $\mathbf{B}$ are antiparallel. In Figure \ref{fig:fliptest}(a), with $\zeta=-20$, the vortices are repelled from the flux tube. This is the opposite of the behaviour seen in Figure \ref{fig:currentalignment}(a), where $\boldsymbol{\Omega}$ and $\mathbf{B}$ are aligned, and similar to the behaviour seen in Figure \ref{fig:currentalignment}(b), where $\boldsymbol{\Omega}$ and $\mathbf{B}$ are aligned but $\zeta$ is positive. The lower panel shows why: vortices move away from the flux tube to produce more aligned current (yellow colouration) near the flux tube. Similarly, in Figures  \ref{fig:currentalignment}(b) and \ref{fig:fliptest}(b) ($\zeta=20$), we see vortex-flux-tube repulsion, when $\boldsymbol{\Omega}$ and $\mathbf{B}$ are aligned, and attraction in the opposite case. Furthermore, in Figures \ref{fig:fliptest}(c) and \ref{fig:fliptest}(d), the plots for $\eta=\pm100$ are identical to the plots for $\eta=\pm100$ in Figure \ref{fig:currentalignment}. In all four bottom panels, the yellow colouration in Figure \ref{fig:currentalignment} is swapped for blue in Figure \ref{fig:fliptest}, because the circulation reverses.

\begin{figure*}
	\includegraphics[width=0.9\textwidth]{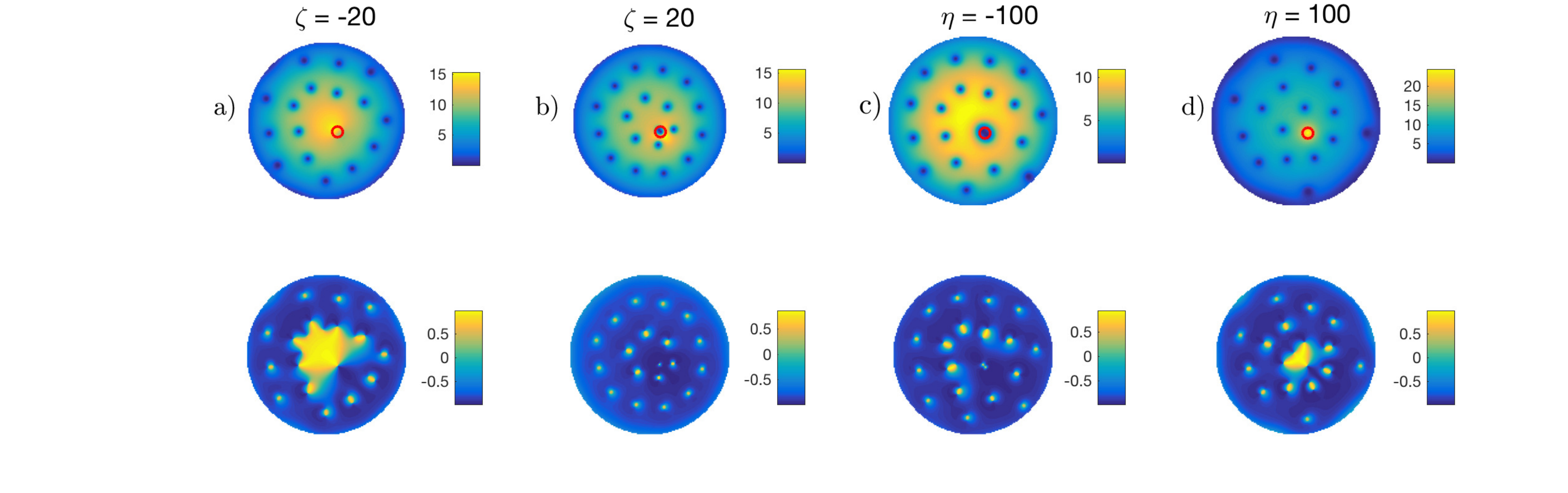}
    \caption{Effect of rotating the magnetic field by $180^{\circ}$ on the attractive and repulsive density and current interactions. The situation is identical to Figure \ref{fig:currentalignment} but with $\mathbf{B}$ and $\boldsymbol{\Omega}$ antialigned. (a) Attractive current coupling $\zeta=-20$. (b) Repulsive current coupling $\zeta=20$. (c) Attractive density coupling $\eta=-100$. (d) Repulsive density coupling $\eta=100$. The top subpanels display the neutron density $|\psi|^2$. The bottom subpanels display the normalised dot product of the current vectors $\mathbf{j}_n \cdot \mathbf{j}_p/j_nj_p$. A colour bar is provided for the contour values in each panel. The units are $n_n$ in the top subpanels. The red circle in the top subpanel marks the position of the off-axis flux tube. Parameters: $\Omega=0.5$, $\tilde{\omega}^2=0.6$, $\tilde{N}_n=N_n/(n_n\xi^{3})=10^3$. In all four bottom panels, the yellow colouration in Figure \ref{fig:currentalignment} is swapped for blue in Figure \ref{fig:fliptest}, because the circulation reverses.}
    \label{fig:fliptest}
\end{figure*}

\subsection{Coupling strength}
\label{subsection:strength}
In Figure \ref{fig:strengthtestminus} we examine the effect on the ground state of increasing the magnitude of $\zeta$. In Figure \ref{fig:strengthtestminus}(a) we see that the coupling is low enough ($\zeta=-0.1$) that $|\psi|^2$ is essentially unchanged compared to the $\zeta=0$ ground state shown in Figure \ref{fig:nocouplingandfluxtube}. Moving  to the right within the figure, as $|\zeta|$ increases from Figure \ref{fig:strengthtestminus}(a) to \ref{fig:strengthtestminus}(e), we see that the vortices move closer to the flux tube (red circle) in the top subpanels and the current becomes increasingly aligned (more yellow colour) in the bottom subpanels. We examine the way the vortex lattice shifts geometrically in response to a coupling to a flux tube coupling in detail in Section \ref{subsection:shift}.

The effect of increasing $|\eta|$ is broadly similar to increasing $|\zeta|$, as demonstrated by Figure \ref{fig:densitystrength_attractive}. For example, vortices tend to overlap with the flux tube in the case of an attractive interaction. We see this for example in Figures \ref{fig:densitystrength_attractive}(d) and \ref{fig:densitystrength_attractive}(e), where two vortices come together and pin at the flux tube location. However, there are differences too. The vortices shift more for $|\zeta|\gtrsim0.1$  in Figure \ref{fig:strengthtestminus} than for $|\eta|\gtrsim1$ in Figure \ref{fig:densitystrength_attractive}. This is not because $|\eta|$ is too low; $|\psi|^2$ does change at the location of the flux tube. The maximum of $\mathbf{j}_p$ occurs at the perimeter of the flux tube, while the minimum of $|\phi|^2$ occurs at the centre, so the response to a density (current-current) coupling is focused at the centre (perimeter). The amount of blue colouration, particularly in panels (c), (d) and (e) of Figures \ref{fig:strengthtestminus} and \ref{fig:densitystrength_attractive}, is less correlated with $|\eta|$ than $|\zeta|$, because density coupling does not favour current alignment between $\mathbf{j}_n$ and  $\mathbf{j}_p$. The changes we see in Figure \ref{fig:densitystrength_attractive} (blue colouration decreases with increasing $|\eta|$) emerge as a side-effect, where pockets of low density overlap with the flux tube.

As the repulsive density coupling increases, we get a mirror image of the behaviour of the attractive coupling for increasing $|\eta|$. For example, while $|\psi|^2$ at the location of the flux tube drops for $\eta<0$, it increases for $\eta>0$. The vortices tend to pin to the flux tube for $\eta<0$ and avoid the flux tube for $\eta>0$. In contrast, the interaction is not always the mirror image when we change from $\zeta<0$ to $\zeta>0$. For example, vortices generally move closer (further) from the flux tube for increasing $|\zeta|$ for $\zeta<0$ ($\zeta>0$), but the vortex patterns are different between $\zeta>0$ and $\zeta<0$. For example, for $\zeta=-0.5$, only the inner ring of vortices shifts, while for $\zeta=0.5$, the outer ring shifts and forms a "front" running through the flux tube position. In addition, $|\psi|^2$ at the flux tube core decreases with increasing $|\zeta|$ for $\zeta>0$ and increases slightly for $\zeta>0$ up to $\zeta=30$, where there is a large drop in density. This happens because a vortex pins to one side of the flux tube for $\zeta\geq30$, even though the current coupling is repulsive.

A vortex pins to one side of the flux tube for large repulsive current-current couplings, because the flux tube is centred off-axis, breaking the rotational symmetry. Hence, vortices prefer to sit on one side of the flux tube rather than the other. We demonstrate this in a schematic in Figure \ref{fig:tikzcircles}(a). There is some non-zero neutron flow on either side of the flux tube, denoted by the purple arrows labelled $\mathbf{n}$ in Figure \ref{fig:tikzcircles}(a). The off-axis flux tube sits in the middle of one of these flows. To the right of the flux tube in Figure \ref{fig:tikzcircles}(a), i.e. on the side nearer the edge of the condensate, $\mathbf{n}$ is aligned with the proton circulation (denoted by the blue arrows marked $\mathbf{p}$). Neighbouring vortices move towards the flux tube to oppose the flow. To the left of the flux tube, nearer the rotation axis, $\mathbf{n}$ and $\mathbf{p}$ are antialigned. If the vortex moves closer to the flux tube, it subtracts from the flow. The logic follows similarly for $\zeta<0$; the vortices prefer to pin on the opposite side of the flux tube to the $\zeta>0$ case. We display this behaviour in action in Figure \ref{fig:tikzcircles}(b): vortices prefer pinning on the bottom right for $\zeta=50$ and on the bottom left for $\zeta=-50$. Importantly, vortices are not equally likely to pin for $\zeta<0$ and $\zeta>0$; pinning for $\zeta>0$ occurs at higher $|\zeta|$ compared to $\zeta<0$.

To study the change in neutron density $|\psi(\mathbf{x}_{\mathrm{FT}})|^2$ at the flux tube core with increasing coupling strength, we plot $|\psi(\mathbf{x}_{\mathrm{FT}})|^2$ versus coupling strength in Figure \ref{fig:densityvalue}. For $\zeta<0$ (top panel), $|\psi(\mathbf{x}_{\mathrm{FT}})|^2$ drops until it reaches almost zero at $|\zeta|=20$ (i.e. one or more vortices pinned at the flux tube location). In contrast, for $\zeta>0$, $|\psi(\mathbf{x}_{\mathrm{FT}})|^2$ increases until $\zeta\geq30$ then drops, when a vortex pins, as discussed above. In Figure \ref{fig:densityvalue}, there is a linear increase (decrease) in $|\psi(\mathbf{x}_{\mathrm{FT}})|^2$ for repulsive (attractive) density couplings. We note that the trends for $\eta>0$ and $\eta<0$ with increasing $|\eta|$ in the bottom panel are mirror images of each other about the horizontal line $|\psi(\mathbf{x}_{\mathrm{FT}})|^2=11.3$.

\begin{figure*}
\centering
\hspace*{-2.3cm} 
	\includegraphics[width=1.2\textwidth]{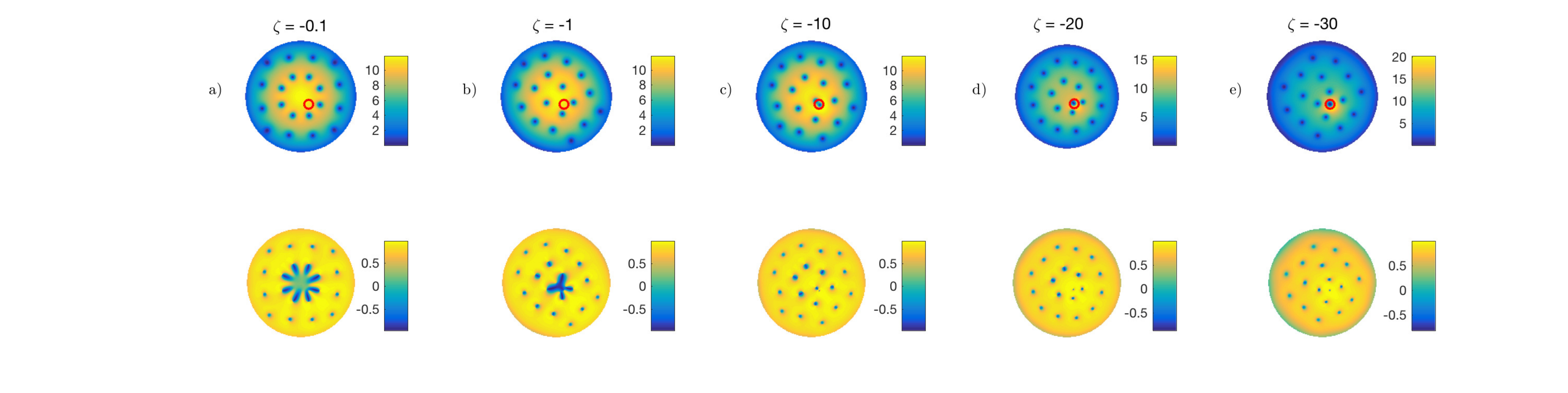}
    \caption{Demonstration of the effect of varying the magnitude of the attractive current coupling, i.e. $\zeta<0$. Ground-state structure of a rotating neutron superfluid coupled to a proton superconductor containing a single off-axis flux tube with $\mathbf{B}$ and $\boldsymbol{\Omega}$ aligned. (a) $\zeta=-0.1$; (b) $\zeta=-1$; (c) $\zeta=-10$; (d) $\zeta=-20$ and (d) $\zeta=-30$. The top subpanels display the neutron density $|\psi|^2$. The bottom subpanels display the normalised dot product of the current vectors $\mathbf{j}_n \cdot \mathbf{j}_p/j_nj_p$. A colour bar is provided for the contour values in each panel. The units are $n_n$ in the top subpanels. The red circle in the top subpanel marks the position of the off-axis flux tube. Parameters: $\Omega=0.5$, $\tilde{\omega}^2=0.6$, $\tilde{N}_n=N_n/(\xi n_n)=10^3$.}
    \label{fig:strengthtestminus}
\end{figure*}

\begin{figure*}
\centering
\hspace*{-2.3cm} 
	\includegraphics[width=1.2\textwidth]{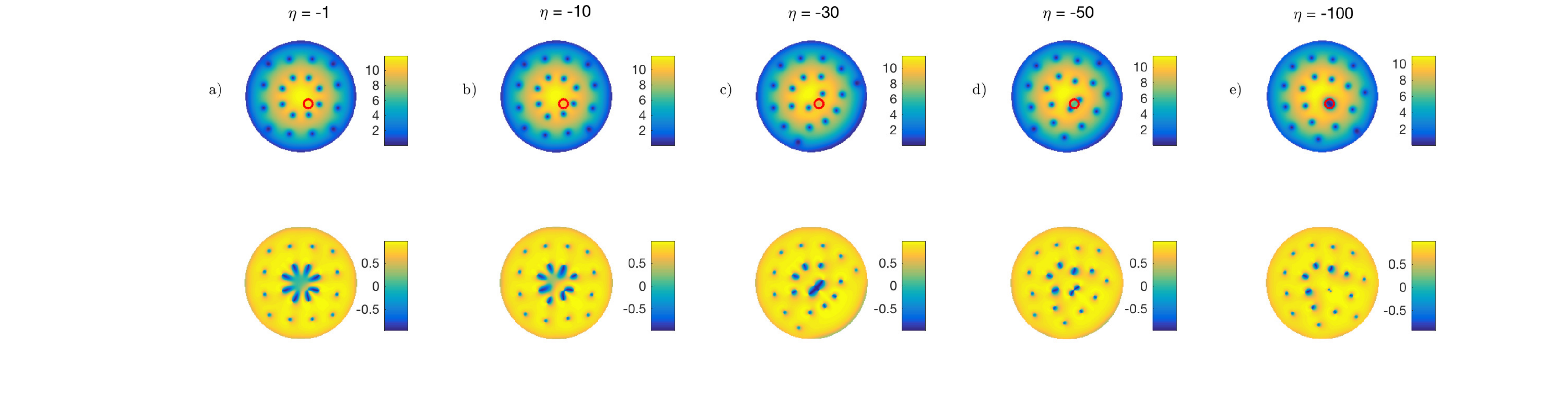}
    \caption{Demonstration of the effect of varying the magnitude of the attractive density coupling, i.e. $\eta<0$. Ground-state structure of a rotating neutron superfluid coupled to a proton superconductor containing a single off-axis flux tube with $\mathbf{B}$ and $\boldsymbol{\Omega}$ aligned. (a) $\eta=-1$; (b) $\eta=-10$; (c) $\eta=-30$; (d) $\eta=-50$ and (d) $\eta=-100$. The top subpanels display the neutron density $|\psi|^2$. The bottom subpanels display the normalised dot product of the current vectors $\mathbf{j}_n \cdot \mathbf{j}_p/j_nj_p$. A colour bar is provided for the contour values in each panel. The units are $n_n$ in the top subpanels. The red circle in the top subpanel marks the position of the off-axis flux tube. Parameters: $\Omega=0.5$, $\tilde{\omega}^2=0.6$, $\tilde{N}_n=N_n/(\xi n_n)=10^3$.}
    \label{fig:densitystrength_attractive}
\end{figure*}

\begin{figure}
\subfloat[]{%
\centering
	\includegraphics[width=1.1\columnwidth]{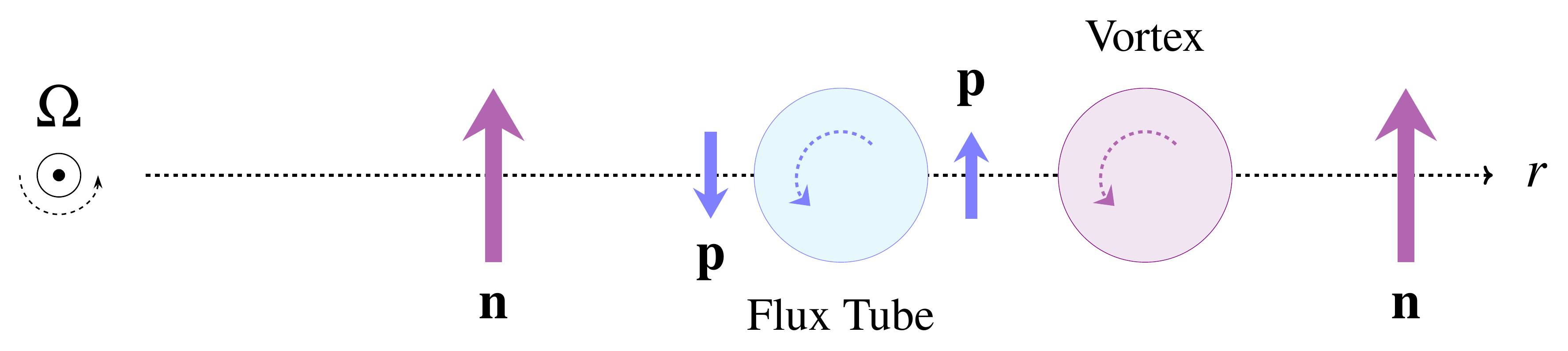}
}

\subfloat[]{%
\centering
\hspace*{-1cm} 
	\includegraphics[width=1.2\columnwidth]{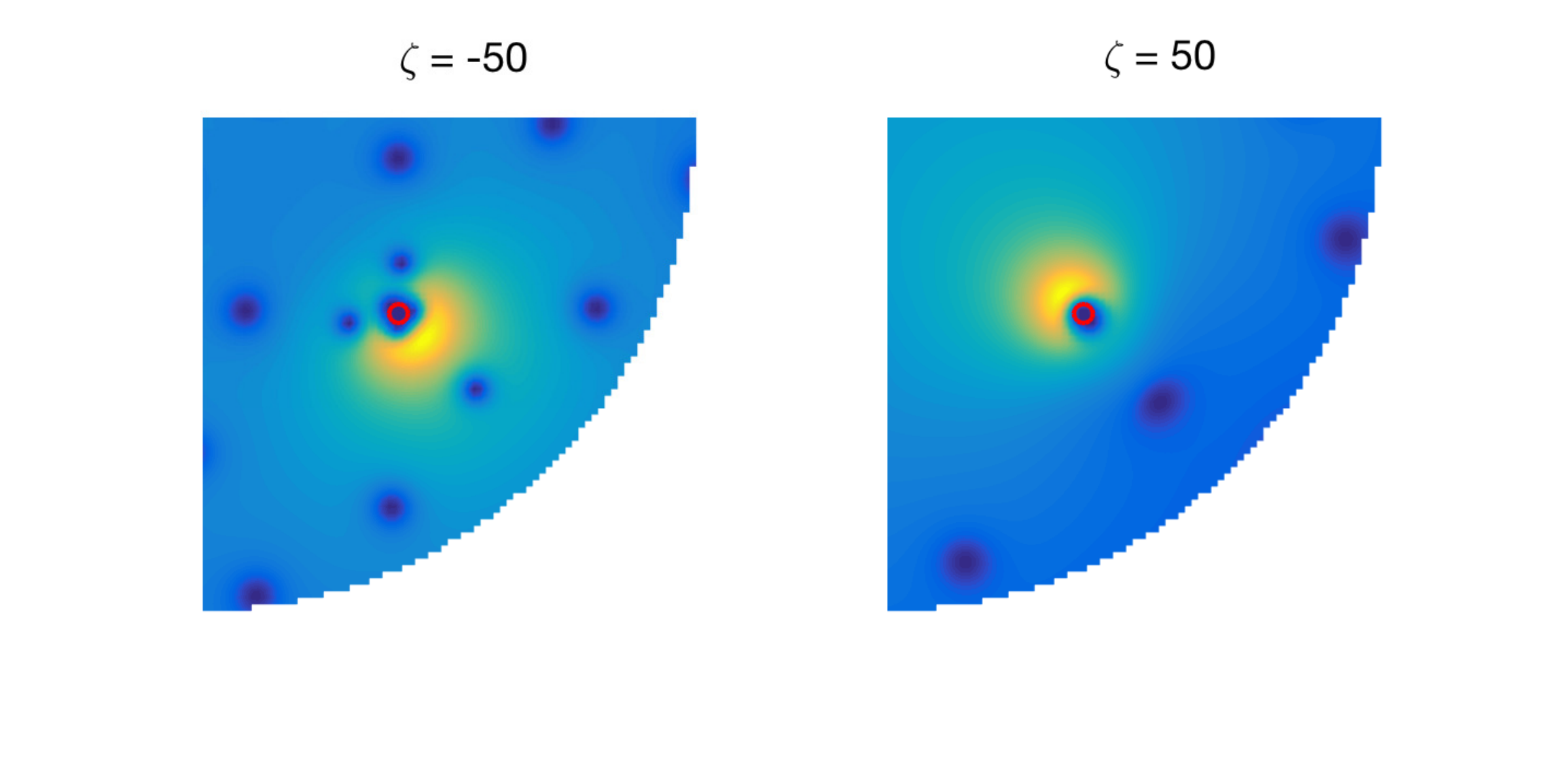}
}

    \caption{(a) Schematic diagram showing how vortices respond to a repulsive current-current coupling with an off-axis flux tube. At the left is the rotation axis, $\boldsymbol{\Omega}$, coming out of the page. The dotted arrow pointing to the right indicates the radial direction $\mathbf{r}$. The global direction of the neutron flow is depicted by the pink arrows labelled $\mathbf{n}$. The blue arrows labelled $\mathbf{p}$ depict the local circulation of the protons around a flux tube (blue circle).  Vortices prefer to sit on one side of the flux tube. (b) Close-up of the neutron density $|\psi|^2$ for $\zeta=-50$ and $\zeta=50$ showing vortices (filled blue circles) pinning to different sides of the flux tube (empty red circle) depending on the sign of the current-current coupling. Parameters: $\Omega=0.45$, $\tilde{\omega}^2=0.3$, $\tilde{N}_n=N_n/(\xi n_n)=4\times10^3$.}
    \label{fig:tikzcircles}
\end{figure}

\begin{figure}
\centering
\includegraphics[width=0.7\columnwidth]{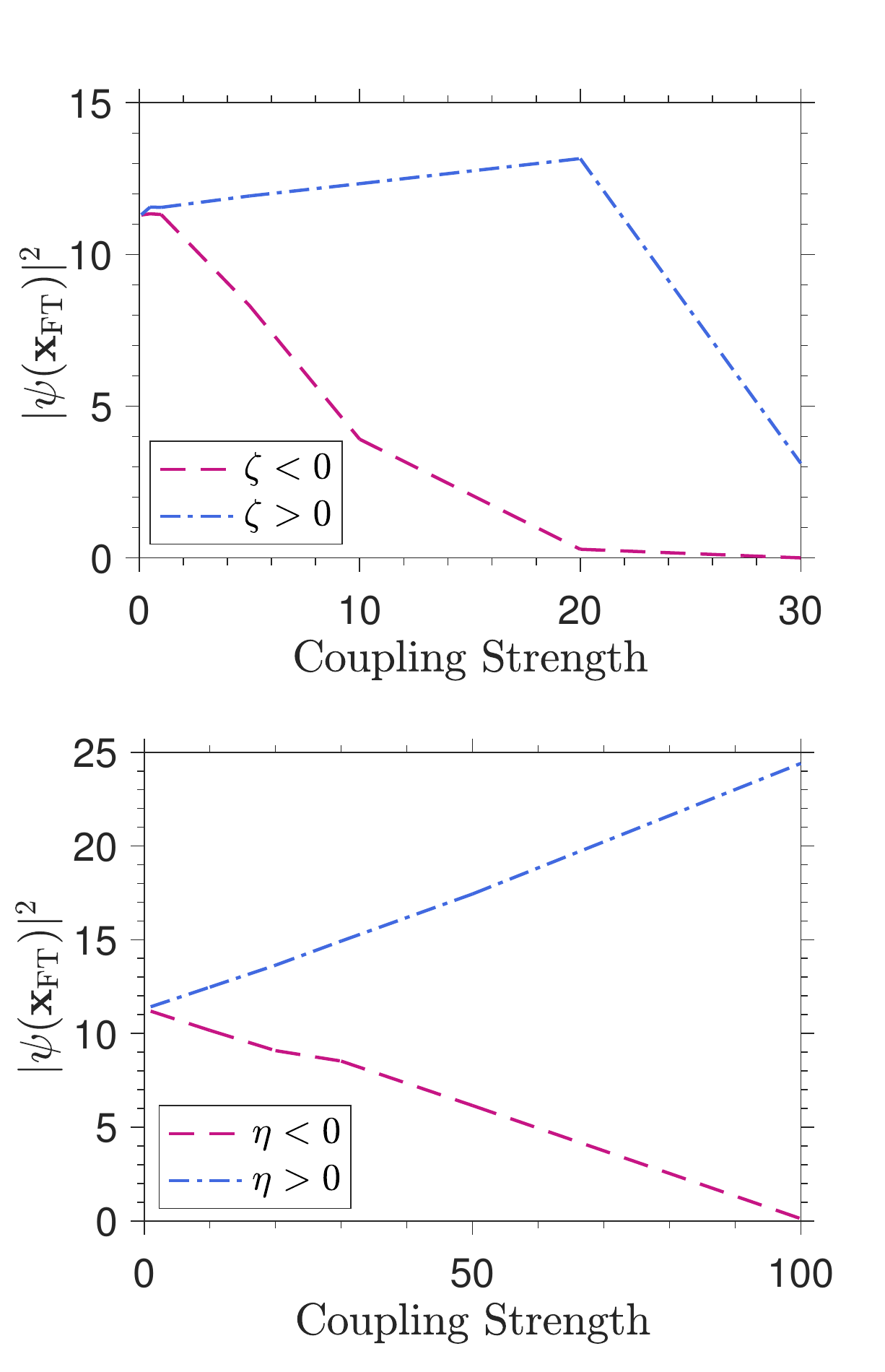}
    \caption{Neutron density $|\psi|^2$ on the flux tube axis, at position $\mathbf{x}_{\mathrm{FT}}$, for a range of coupling strengths. Top panel: $|\psi(\mathbf{x}_{\mathrm{FT}})|^2$ versus $|\zeta|$ for $\zeta<0$ (pink curve) and $\zeta>0$ (blue curve). Bottom panel: $|\psi(\mathbf{x}_{\mathrm{FT}})|^2$ versus $|\eta|$ for $\eta<0$ (pink curve) and $\eta>0$ (blue curve). $|\psi(\mathbf{x}_{\mathrm{FT}})|^2$ is in units of $n_n$. Parameters: $\Omega=0.5$, $\tilde{\omega}^2=0.6$, $\tilde{N}_n=N_n/(\xi n_n)=10^3$.}
    \label{fig:densityvalue}
\end{figure}

\subsection{Abrikosov lattice rearrangement}
\label{subsection:shift}
We define a metric to quantify how the lattice as a whole reorganises in response to the flux tube coupling: $\langle d_\mathrm{FT}^2 \rangle$, the mean-square distance of each vortex from the flux tube. In this section, we study a larger system ($40\times40$ box with 60--80 vortices) than in Sections \ref{subsection:attractvsrepulse}--\ref{subsection:strength}. We include in the metric the 20 vortices nearest to the flux tube for consistency. \footnote{The condensate boundary changes with $\eta$ and $\zeta$, as does the number of vortices it encloses. If one is not careful, this obscures the behaviour we are trying to analyse, i.e. the response of the vortex lattice as a whole to the flux tube coupling. All our simulations contain $\geq20$ vortices at all times.}

Consider the following experiment. Let us introduce a flux tube into the neutron condensate with an attractive interaction ($\eta<0$ or $\zeta<0$). Suppose the closest vortex pins to the flux tube. The remaining vortices experience competing forces: it is favourable for them to overlap with the flux tube, but they are also repelled by the vortex now at that position. What happens? Consider first the top row of Figure \ref{fig:distancemetrics_sq} (current coupling). For a repulsive interaction, we expect $\langle d_\mathrm{FT}^2 \rangle$ to grow with $|\zeta|$, if the vortices move away from the flux tube and vice versa for an attractive interaction. Therefore, the curves diverge as the vortices either pile on top of or move away from the flux tube. When we exclude vortices pinned to the flux tube (within $3\xi_p$ of the flux tube axis), the repulsive current interaction causes $\langle d_\mathrm{FT}^2 \rangle$ to continue to rise past $\zeta>10$ [panel (b)]. When we include the pinned vortices [panel (a)], $\langle d_\mathrm{FT}^2 \rangle$ drops for $\zeta>10$, because vortices pin to one side of the flux tube (see Figure \ref{fig:tikzcircles}), while the unpinned vortices move further away with increasing $\zeta$. For $\zeta<0$, we see a big drop in $\langle d_\mathrm{FT}^2 \rangle$ when we include the pinned vortices in panel (a) but very little change when we exclude them [panel (b)], indicating that $\langle d_\mathrm{FT}^2 \rangle$ drops because vortices pile on top of the flux tube.

Now consider the bottom row of Figure \ref{fig:distancemetrics_sq} (density coupling). When the pinned vortices are included, $\langle d_\mathrm{FT}^2 \rangle$ doesn't change very much. When the pinned vortices are excluded, the behaviour for $\eta>0$ and $\eta<0$ is similar for $|\eta|<100$ and the vortices shift slightly. The divergent behaviour for $|\eta|>100$, on the other hand, seems counter-intuitive: $\langle d_\mathrm{FT}^2 \rangle$ increases with $|\eta|$ for $\eta>0$. One interpretation is that the attractive density interaction favours vortices pinning directly on top of the flux tube but does not affect where the other mutually repelling vortices end up. Similarly, the repulsive density interaction concentrates as much neutron density as possible onto the flux tube location but does not influence the other vortices much. In Figure \ref{fig:densityvalue}, we see that $|\psi|^2$ on the flux tube axis increases from $11.4n_n$ to $24.4n_n$, as $|\eta|$ increases from $1$ to $100$, but otherwise the mutually repelling vortices arrange themselves without much regard for the flux tube. The mutual vortex repulsion (irrespective of the sign of $\eta$) explains why, for $\eta$ large and negative, vortices that do not sit directly on top of the flux tube move further away, while, for $\eta$ large and positive, there is no vortex sitting on top of the flux tube, so the vortices cluster closer together.

We remind the reader that flux tubes are more numerous than vortices in a neutron star ($n_{\Phi}>n_{v}$). The trends in Figure \ref{fig:distancemetrics_sq} are instructive for understanding the local interaction physics, but more work is needed to understand fully the global rearrangement of a vortex lattice under neutron star conditions.

The density and current couplings depend differently on the order parameters. For the density coupling, all that matters is $|\psi|^2|\phi|^2$. For $\eta>0$, for example, the neutron vortices become shallower (heal to a lower background density), as neutrons drain from the rest of the condensate to the flux tube's position, creating a mountain in $|\psi|^2$ at that spot and a valley everywhere else with  "divots" at each vortex.  In contrast, the current coupling depends on the magnitude and direction of $\mathbf{j}_n$ and $\mathbf{j}_p$ (Figure \ref{fig:strengthtestminus}). For example, for $\zeta>0$, the vector interaction operates such that regions of large $\mathbf{j}_n$ overlap with regions of large $\mathbf{j}_p$, and $\mathbf{j}_n$ and $\mathbf{j}_p$ are opposed. The higher the current locally, the more important it is energetically for $\mathbf{j}_n$ and $\mathbf{j}_p$ to be opposed at that point. Simply draining the neutrons  towards the flux tube is not enough, unlike for the density coupling. The currents are redirected, i.e. the quantised vortices must move away from the flux tube to avoid having $\mathbf{j}_n$ parallel to $\mathbf{j}_p$.

\begin{figure*}
\centering
\includegraphics[width=0.65\textwidth]{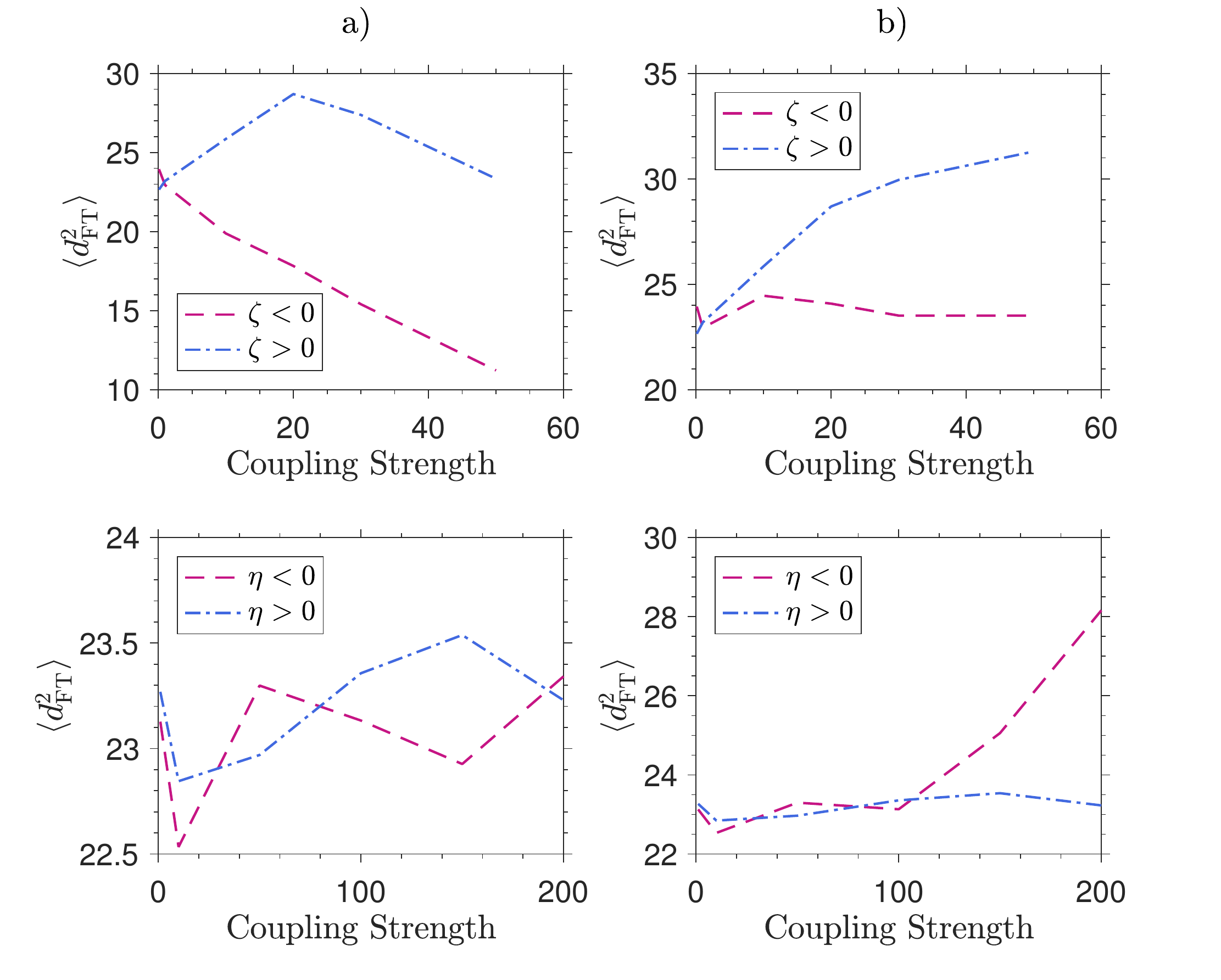}
    \caption{Vortex lattice rearrangement: spacing metric $\langle d_\mathrm{FT}^2 \rangle$ (units of $\xi_n^2$) versus coupling strength. Top panels: $\langle d_\mathrm{FT}^2 \rangle$ versus $|\zeta|$ for $\zeta<0$ (pink curve) and $\zeta>0$ (blue curve). Bottom panels: $\langle d_\mathrm{FT}^2 \rangle$ versus $|\eta|$ for $\eta<0$ (pink curve) and $\eta>0$ (blue curve). (a) Vortices pinned to the flux tube are included. (b) Vortices pinned to the flux tube are excluded. Parameters: $\Omega=0.45$, $\tilde{\omega}^2=0.3$, $\tilde{N}_n=N_n/(\xi n_n)=4\times10^3$.}
    \label{fig:distancemetrics_sq}
\end{figure*}

\subsection{Energetics}
\label{subsection:energy}
\begin{figure*}
	\includegraphics[width=0.7\textwidth]{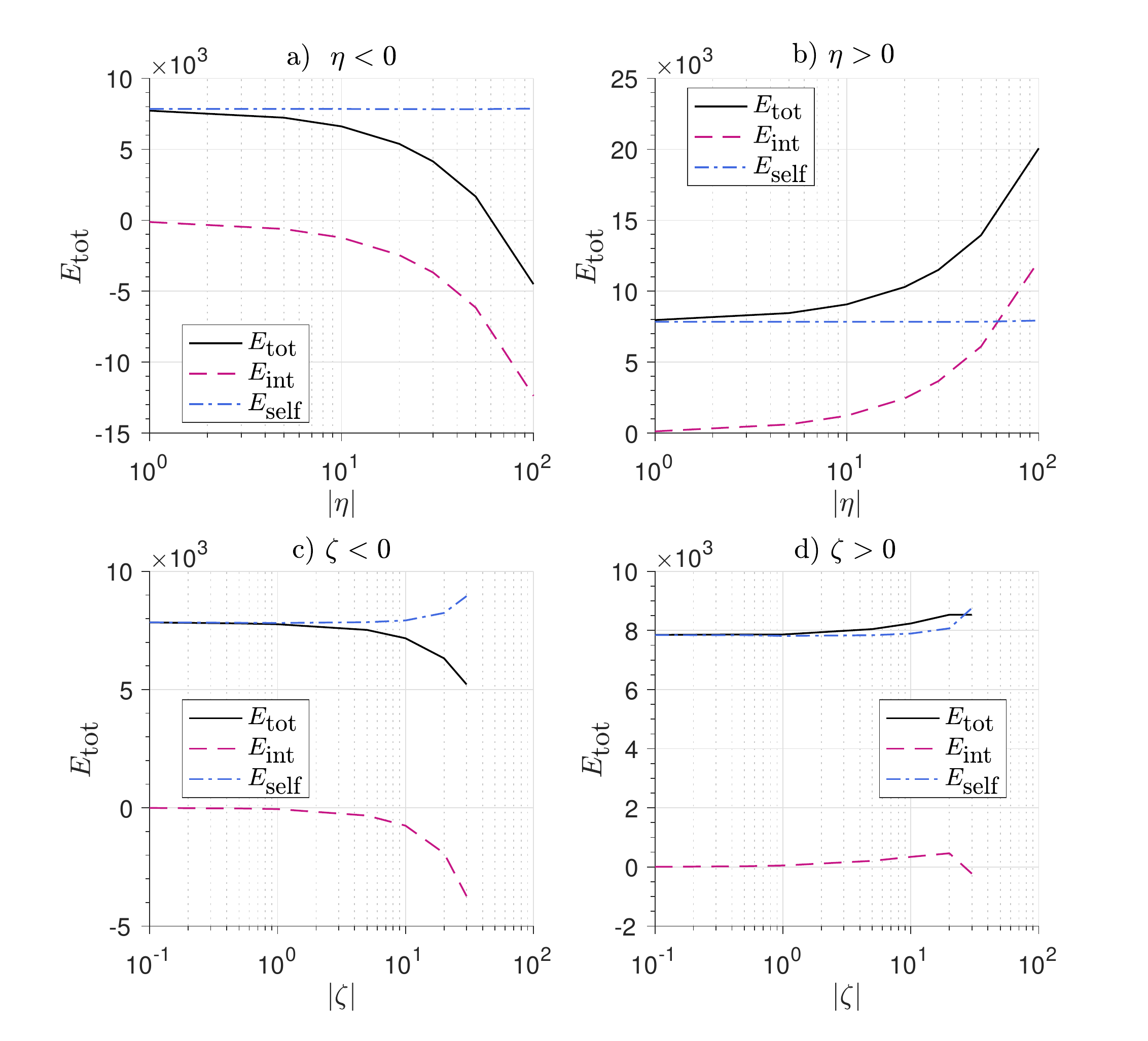}
    \caption{Kinetic and potential energy $E_{\mathrm{tot}}$ (in units of $n_nU_0$) for a range of coupling strengths: total energy $E_{\mathrm{tot}}$ (black curve), interaction energy $E_{\mathrm{int}}$ (pink curve) and self energy $E_{\mathrm{self}}$ (blue curve). (a) $\eta<0$. (b) $\eta>0$. (c) $\zeta<0$. (d) $\zeta>0$. Parameters: $\Omega=0.5$, $\tilde{\omega}^2=0.6$, $\tilde{N}_n=N_n/(\xi n_n)=10^3$.}
    \label{fig:energyplot}
\end{figure*}

The dimensionless total energy of the neutron condensate in the rotating frame is given by $E_{\mathrm{tot}}=E_{\mathrm{self}}+E_{\mathrm{int}}$, where
\begin{equation}
E_{\mathrm{self}}[\psi]=\int \mbox{d}^3 \mathbf{x} \ \left( \frac{1}{2}|\nabla\psi|^2+V|\psi|^2+\frac{1}{2}|\psi|^4-\Omega\psi^*\hat{L}_z\psi  \right)
\end{equation}
adds the kinetic energy to the contributions from the trap, boson self-attraction and rotation, and $E_{\mathrm{int}}$ is associated with the neutron-proton coupling $\mathcal{H}_{int}[\psi,\phi]$, defined in equation (\ref{eq:Eint1}). It is important to recognise that $E_{\mathrm{self}}$ is not the total energy of the ground state in the absence of neutron-proton coupling as in Figure \ref{fig:nocouplingandfluxtube}. Rather it is the self-interacting part of the energy budget for the ground state when the neutron-proton coupling is switched on; $\mathcal{H}_{int}$ affects the structure of $\psi$ and hence $E_{\mathrm{self}}$ as well as $E_{\mathrm{int}}$.

We calculate $E_{\mathrm{tot}}$, $E_{\mathrm{self}}$ and $E_{\mathrm{int}}$ for a range of density and current coupling strengths in Figure \ref{fig:energyplot}. We find that $E_{\mathrm{self}}$ is broadly (but not completely) independent of $|\eta|$ and $|\zeta|$. Hence, $E_{\mathrm{int}}$ dominates the trends seen in $E_{\mathrm{tot}}$. However, $E_{\mathrm{self}}$ does depend weakly on $|\eta|$ and $|\zeta|$, because the coupling modifies $\psi$ and its gradients, especially where the vortices end up standing. For example, in Figures \ref{fig:energyplot}(c) and \ref{fig:energyplot}(d), we see that $E_{\mathrm{self}}$ changes significantly compared to the zero coupling case for $\zeta=20$ and $\zeta=30$.  We emphasise again that the domains of $\eta$ and $\zeta$ under consideration are chosen to illustrate the physics rather than model a realistic neutron star, although they are broadly consistent with (\ref{eq:eta}) and (\ref{eq:zeta}).

Equation (\ref{eq:Eint1}) implies that $E_{\mathrm{int}}$ decreases with $|\eta|$ and $|\zeta|$ for $\eta$, $\zeta<0$ and increases for $\eta$, $\zeta>0$, which is what we observe. However, the current coupling, which depends on orientation (see Section \ref{subsection:orientation}), behaves differently. When we compare the situations $\eta<0$ and $\eta>0$, they are symmetric, with $\eta<0$ ($\eta>0$) producing a decrease (increase) in $E_{\mathrm{tot}}$ of $158\%$ over the range $\eta=1$--$10^2$. In contrast, the magnitude of the change in energy compared to zero coupling for $\zeta>0$ is smaller ($8.68\%$ increase) than for $\zeta<0$ ($33.4\%$ decrease) over the range $\zeta=0.1$--$30$. This is because $\zeta>0$ favours anti-alignment of the currents; there are more regions of negative $\mathbf{j}_n \cdot \mathbf{j}_p$ contributing to $E_\mathrm{int}$, which reduces  $E_\mathrm{int}$. In fact, for $\zeta=30$, there is enough antialigned current that $E_\mathrm{int}$ is negative. \citet{AlfordGood2008} observed a similar asymmetry in their one-dimensional calculation.

\section{Flux Tube Array}
\label{section:FTA}
We now generalise the single-flux-tube studies in Section \ref{section:SFT} to an array of flux tubes and simultaneously step up from two to three dimensions. This situation approximates more closely what happens in the outer core of a neutron star, although we emphasize again that the simulation volume and parameters lie well outside the neutron star regime due to computational limitations. In Section \ref{subsection:VT} we present a typical ground state generated by the simulations, which exemplifies an essential property: the neutron vortices form a "tangle", i.e. a frustrated equilibrium which represents a compromise between competing interactions discussed in Section \ref{subsection:glassy}. In Section \ref{subsection:orientation3D}, we study how the structure of the vortex tangle depends on the angle between $\boldsymbol{\Omega}$ and $\mathbf{B}$.

\subsection{Vortex tangle}
\label{subsection:VT}

\begin{figure}
	\includegraphics[width=1\columnwidth]{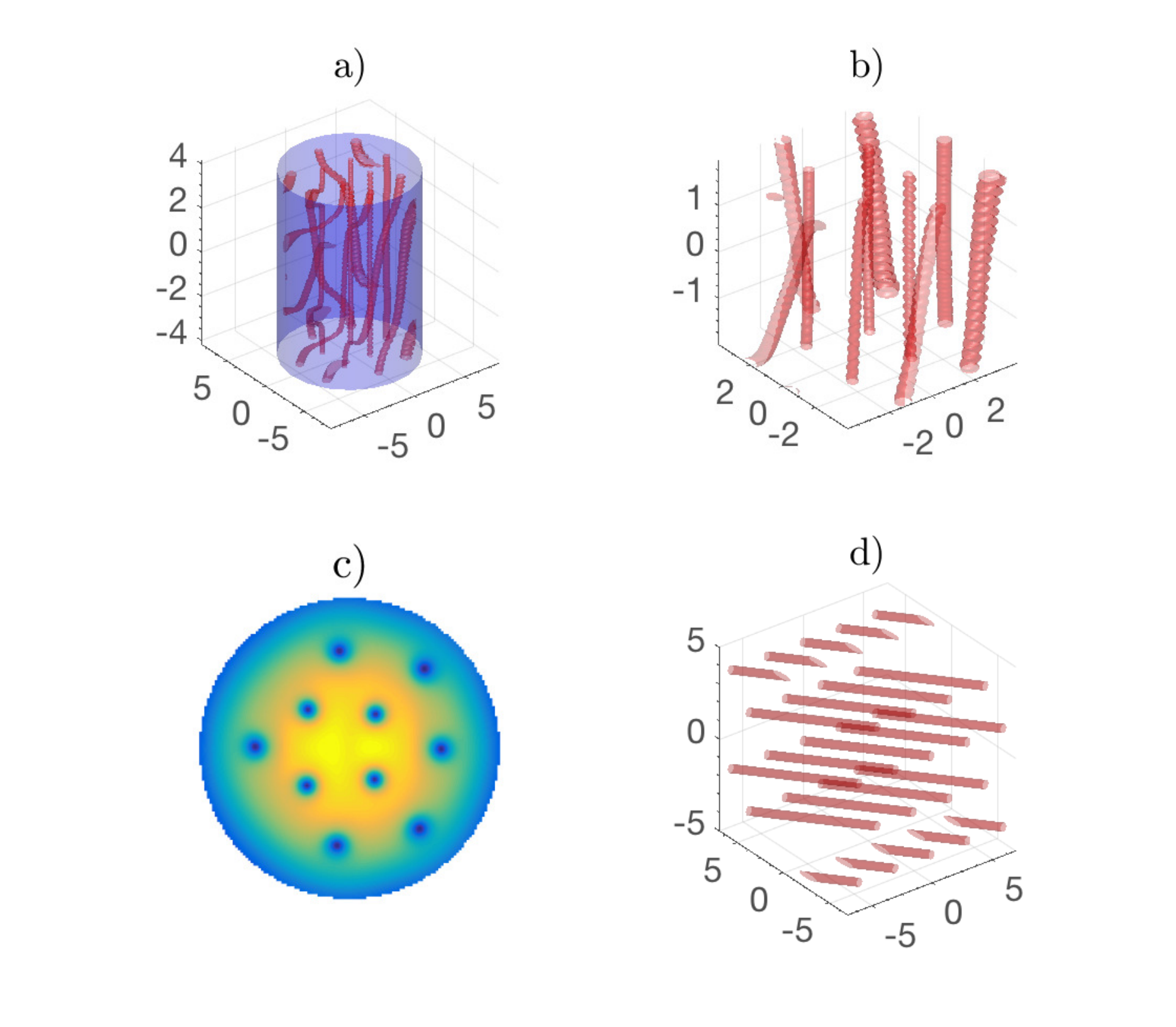}
    \caption{Structure of a three-dimensional ground state featuring a vortex tangle. (a) Vortex structure. Blue shading denotes the condensate volume (drawn at the furthest distance from the axis, where $|\psi|^2$ drops below $10\%$ of its maximum). Red shading traces out vortices (drawn where $|\psi|^2$ drops below $10\%$ of its maximum inside the vortex core). (b) Close up vortices in (a). (c) Cross-section of $|\psi|^2$ (in units of $n_n$) through $z=0$. (d) Flux tube array. Red shading traces out vortices (drawn where $|\phi|^2$ drops below $10\%$ of its maximum inside the flux tube core). Parameters: $\tilde{N}_n=8000$, $\Omega=0.5$, $\tilde{N}_p=0.05\times \tilde{N}_n$, $d_{\Phi}=3$.}
    \label{fig:repeg}
\end{figure}

\begin{figure}
	\includegraphics[width=\columnwidth]{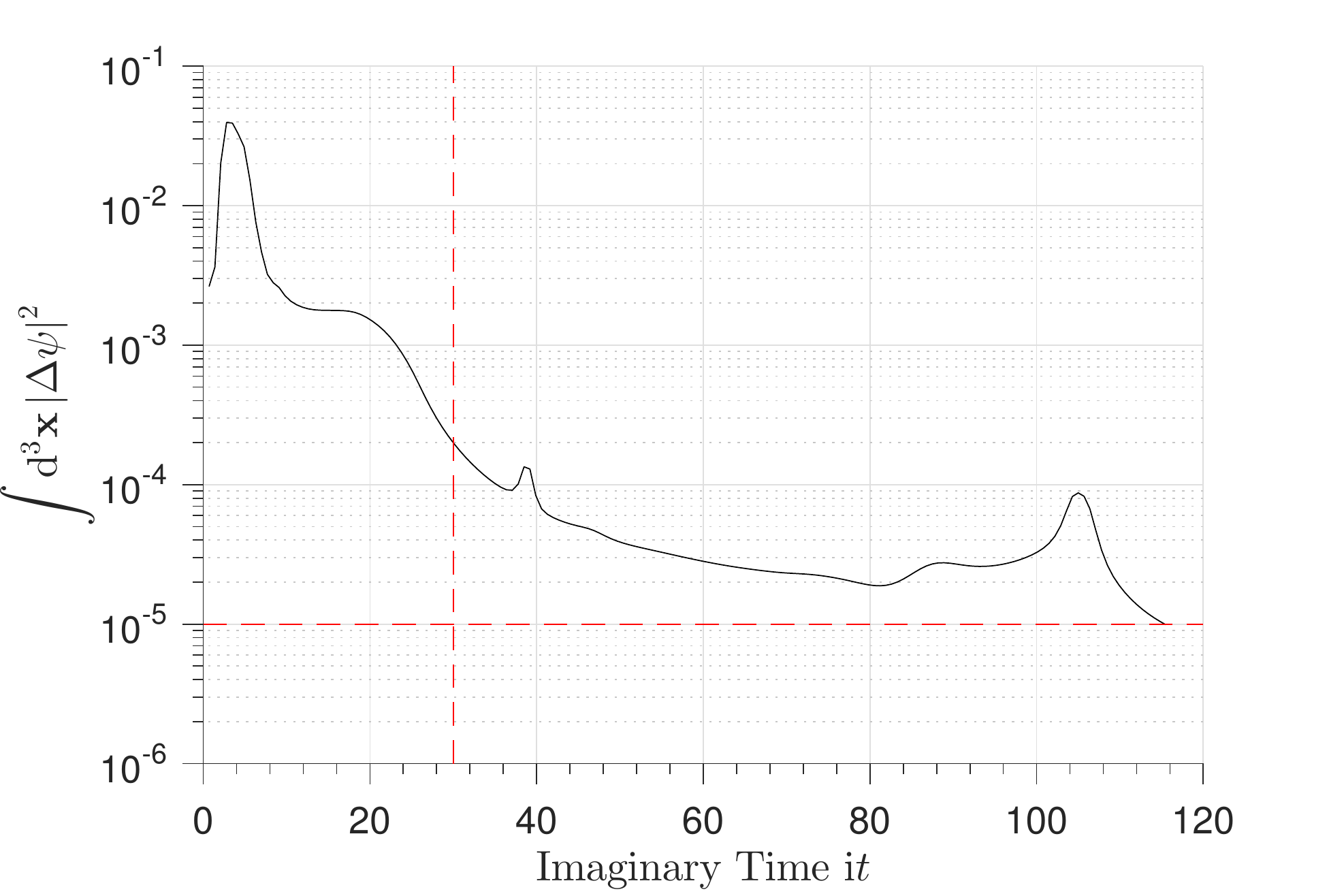}
    \caption{Convergence metric $\int \mbox{d}^3 \mathbf{x} \ |\Delta \psi|^2$ (in units of $N_n$) versus imaginary time $\mathrm{i}t$ (in units of $\xi_n/c_s$) for the state in Figure \ref{fig:repeg}. $\Delta \psi$ denotes the difference between $\psi$ at successive imaginary time-steps. Red dashed lines signify convergence thresholds (see text for definition).}
    \label{fig:convplot}
\end{figure}

Figure \ref{fig:repeg} displays a representative example of a three-dimensional ground state, where the neutrons couple to the protons via a density coupling with $\eta=-10$ and $\theta=75^{\circ}$.  The figure visualises the neutron fluid in three ways. Figure \ref{fig:repeg}(a) displays surfaces (in red) of $|\psi|^2$ drawn at $10\ \%$ of the maximum, showing vortices as curved tubes. Figure \ref{fig:repeg}(b) is a close up of Figure \ref{fig:repeg}(a), while Figure \ref{fig:repeg}(c) shows a cross-section of $|\psi|^2$ through the midplane $z=0$. The flux tubes are depicted in Figure \ref{fig:repeg}(d), where a surface of $|\phi|^2$ (in red) is drawn at $10\ \%$ of the maximum value. The flux tubes have $\theta=75^{\circ}$ and are placed in a triangular Abrikosov lattice. As in a neutron star, we have $d_v\geq d_{\Phi}$, with $\sim20$ flux tubes and $\sim10$ vortices in the condensate (note: $d_v\gg d_{\Phi}$ in a neutron star). We plot 20 flux tubes in Figure \ref{fig:repeg} to avoid overcrowding. In later simulations (e.g. Figures \ref{fig:thetaplotwithinsets_small2} and \ref{fig:linemeasures_plot}) we have $\sim50$ flux tubes and $\sim10$ vortices.

We use a vortex finding algorithm to identify the vortex intersection in each plane $z=\mbox{constant}$ by locating velocity maxima [\citet{Jamesprivate}; see also \citet{Douglass2015}]. We connect nearest neighbours in adjacent planes and locate the terminus, where no neighbour exists within a distance threshold. Once the ordering of the points along the vortex cores is established, we identify contiguous filaments using a depth-first-search algorithm and smoothing spline function.\footnote{We use the smoothing spline MATLAB function \texttt{spaps} from the Curve Fitting Toolbox to return the smoothest function within a $5\times10^{-3}$ tolerance. Refer to the MATLAB documentation for more details.} Letting $\mathbf{s}(\xi)$ denote the displacement of an arbitrary point on the filament from the origin, with affine parameter $0\leq\xi\leq1$, we define two global properties of a vortex tangle:  $\langle \kappa \rangle$, the curvature averaged over all the vortices, and $L$, the total vortex length:
\begin{equation} 
\label{eq:length}
L=\int \mbox{d} \xi \ |\mathbf{s'(\xi)}|  ,
\end{equation}
\begin{equation}
 \label{eq:curvature}
\langle \kappa \rangle=\frac{1}{L}\int \mbox{d} \xi \ \frac{|\mathbf{s'(\xi)}| | \mathbf{s}'(\xi)\times \mathbf{s}''(\xi)|}{|\mathbf{s}'(\xi)|^3}   .
\end{equation}
In (\ref{eq:length}) and (\ref{eq:curvature}), a prime denotes differentiation with respect to $\xi$. Vortex length per unit volume is a fundamental global property of a vortex tangle and captures many features of superfluid turbulence \citep{donnelly1991}. Mutual friction and vortex reconnection drive the growth and decay of $L$ in superfluid turbulence, e.g. in liquid helium \citep{Barenghi2001}. The reciprocal mean curvature $\langle \kappa \rangle^{-1}$ quantifies the mean radius of curvature on which the vortices wrinkle in the tangle.

The state pictured in Figure \ref{fig:repeg} has $L=137.9$ and $\langle\kappa\rangle=0.352$. The corresponding $\theta=0^{\circ}$ state with all other parameters identical has 12 straight vortices each $10$ units long, i.e. $L=120.0$  and $\langle\kappa\rangle=0$.

 \subsection{Glassiness and frustration}
 \label{subsection:glassy}
 When the neutron-proton coupling is weak ($|\eta|\lesssim1$, $|\zeta|\lesssim0.1$), the vortex and flux tube arrays configure independently. When the coupling is strong, the arrays pin perfectly to one another. When the coupling is intermediate, interesting and complicated behaviour arises due to the competition between vortex-flux-tube interactions (which may be attractive or repulsive) and vortex-vortex interactions (which are always repulsive, if the circulations of both vortices are in the same sense). This competition leads to glassy relaxation in imaginary time: many metastable configurations exist with different layouts yet similar energies, so the system takes a long time (and many "false starts") to navigate the complicated energy landscape and find the ground state.
 
 In order to study the system's glassiness, we define a convergence metric $\int \mbox{d}^3 \mathbf{x} \ |\Delta \psi|^2$, where $\Delta \psi$ is the difference in the wavefunction between successive imaginary time steps. We choose $\int \mbox{d}^3 \mathbf{x} \ |\Delta \psi|^2 \leq10^{-5} N_n $ as the arbitrary numerical tolerance, where we stop imaginary time propagation and declare that convergence is achieved.  Additionally, we demand $t\geq30$ in order to consider the state converged, for reasons explained below. Figure \ref{fig:convplot} shows how  $\int \mbox{d}^3 \mathbf{x} \ |\Delta \psi|^2$ evolves with imaginary time in a typical simulation. Convergence occurs when the curve drops below the dashed horizontal line to the right of the dashed vertical line. We introduce the $t\geq30$ constraint from experience gained by observing multiple simulations. Large fluctuations in the convergence metric for $t<30$ can take it below $10^{-5}N_n$, temporarily, even though the ground-state is not yet reached. Fortunately, the "fickleness" of the system decreases as imaginary time passes: the "quakes" (vortex reconfigurations) that pave the way to the next metastable state occur increasingly far apart \citep{Anderson2004}. The $t\geq30$ constraint allows the system to evolve away from the arbitrary choice of initial wavefunction, skip past the metastable states it finds initially, and explore the landscape of lowest-energy configurations. If the state happens to truly converge by luck at $t<30$, it remains converged at $t\geq30$, and we pay no penalty except for a minor increase in computational cost.
 
 We study the progression through metastable states in more detail in Figure \ref{fig:glassyfigure}. The figure displays the convergence metric and $E_{\mathrm{tot}} $ versus imaginary time in the top two panels. We see spikes or "quakes", which correspond to sudden reconfigurations. They cause the convergence metric to rise temporarily but their cumulative effect is to lead the system towards greater stability; the convergence metric drifts downwards in the longer term. The lower two rows of panels show $|\psi|^2$ (contour plot for cross-section through midplane; surface plot in three dimensions) for intermediate states near the quakes. States (a) and (b) at $t\leq30$ retain some memory of the initial state. In state (c) the "ghost vortices" that form at the edge of the condensate move inward; the state is still not converged by our criterion, i.e. we still have $\int \mbox{d}^3 \mathbf{x} \ |\Delta \psi|^2 >10^{-5} N_n $. In the bottom row of three-dimensional plots, we see that the configuration changes differently in each plane $z=\mbox{constant}$, as imaginary time elapses. 
 
Glassy systems involve frustration, where a unique minimum energy state state is hard or impossible to find due to the conflict between different interactions. The flux tube array tilted at an angle relative to $\boldsymbol{\Omega}$ introduces more frustration into the system. The potential for topological complexity in three dimensions means that vortex tangles form, as the system relaxes. Does access to these plentiful, frustrated, geometrically complicated states in three dimensions open up an abundance of new dynamics, when the system is driven? We will explore this open question in a future article.

\begin{figure*}
	\includegraphics[width=\textwidth]{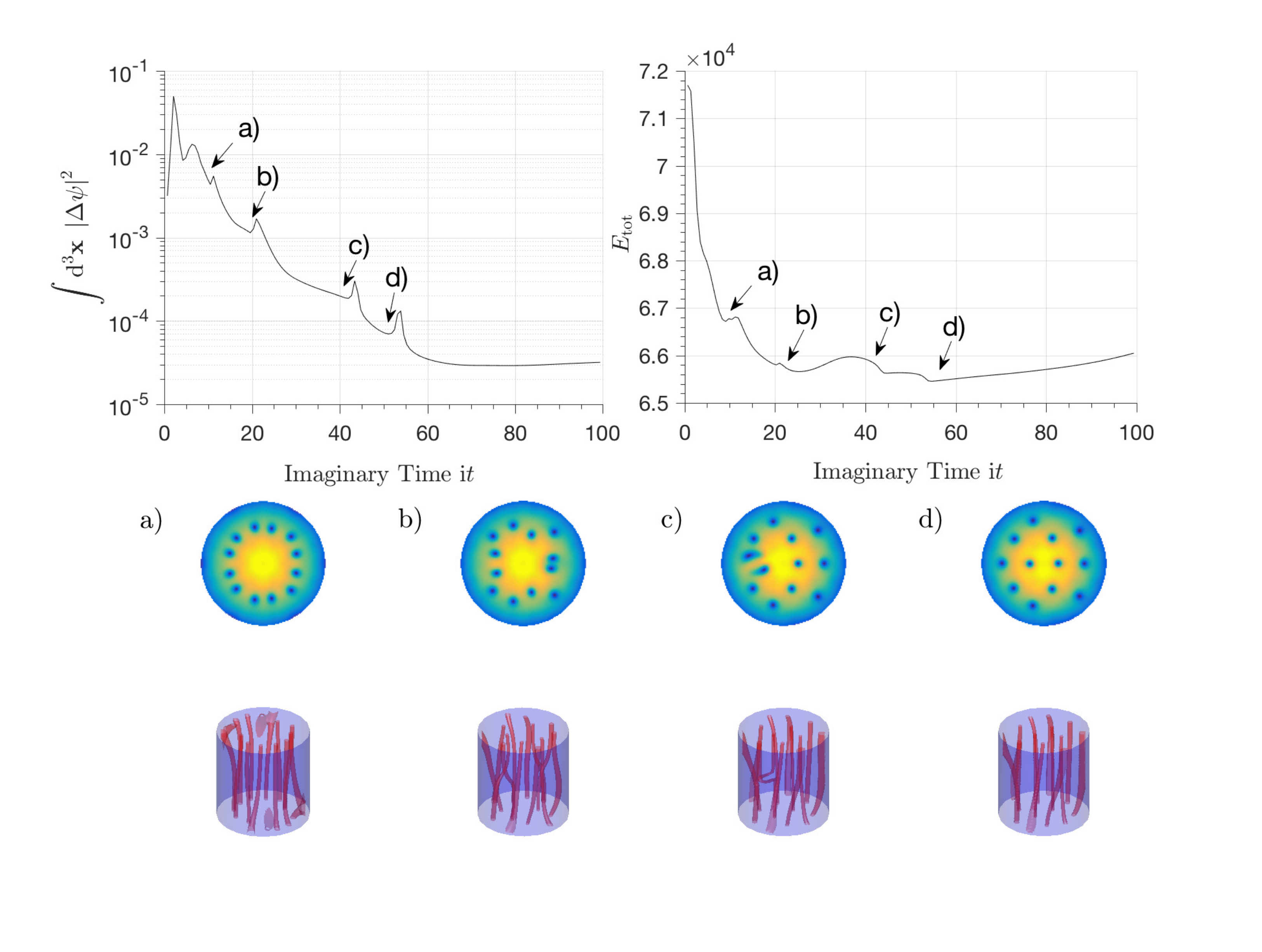}
    \caption{Convergence of a glassy system of neutron vortices coupled to proton flux tubes tilted at $\theta=10^{\circ}$ relative to the rotation axis $\boldsymbol{\Omega}$ (density coupling $\eta=-10$). (\textit{Top}) Convergence metric $\int \mbox{d}^3 \mathbf{x} \ |\Delta \psi|^2$ (in units of $N_n$) and total energy $E_{\mathrm{tot}}$ (in units of $n_nU_0$) versus imaginary time (in units of $\xi_n/c_s$). (\textit{Middle}) Cross-sections of $|\psi|^2$ through $z=0$ (arbitrary units). (\textit{Bottom}) Three-dimensional surface plots of $|\psi|^2$ (arbitrary units), where the red shading traces out vortices (drawn where $|\psi|^2$ drops below $10\%$ of its maximum in the core) and the blue shading is the condensate edge. Panels (a)--(d) correspond to snapshots of the convergence indicated by arrows in the top panels. Parameters: $\tilde{N}_n=8\times10^3$, $\Omega=0.5$, $\tilde{N}_p=0.05\times \tilde{N}_n$, $d_{\Phi}=3$.}
    \label{fig:glassyfigure}
\end{figure*}

\subsection{Relative orientation}
\label{subsection:orientation3D}
How do the vortex tangles computed in Sections \ref{subsection:VT} and \ref{subsection:glassy} depend on the angle between $\boldsymbol{\Omega}$ and $\mathbf{B}$?

\begin{figure*}
	\includegraphics[width=0.85\textwidth]{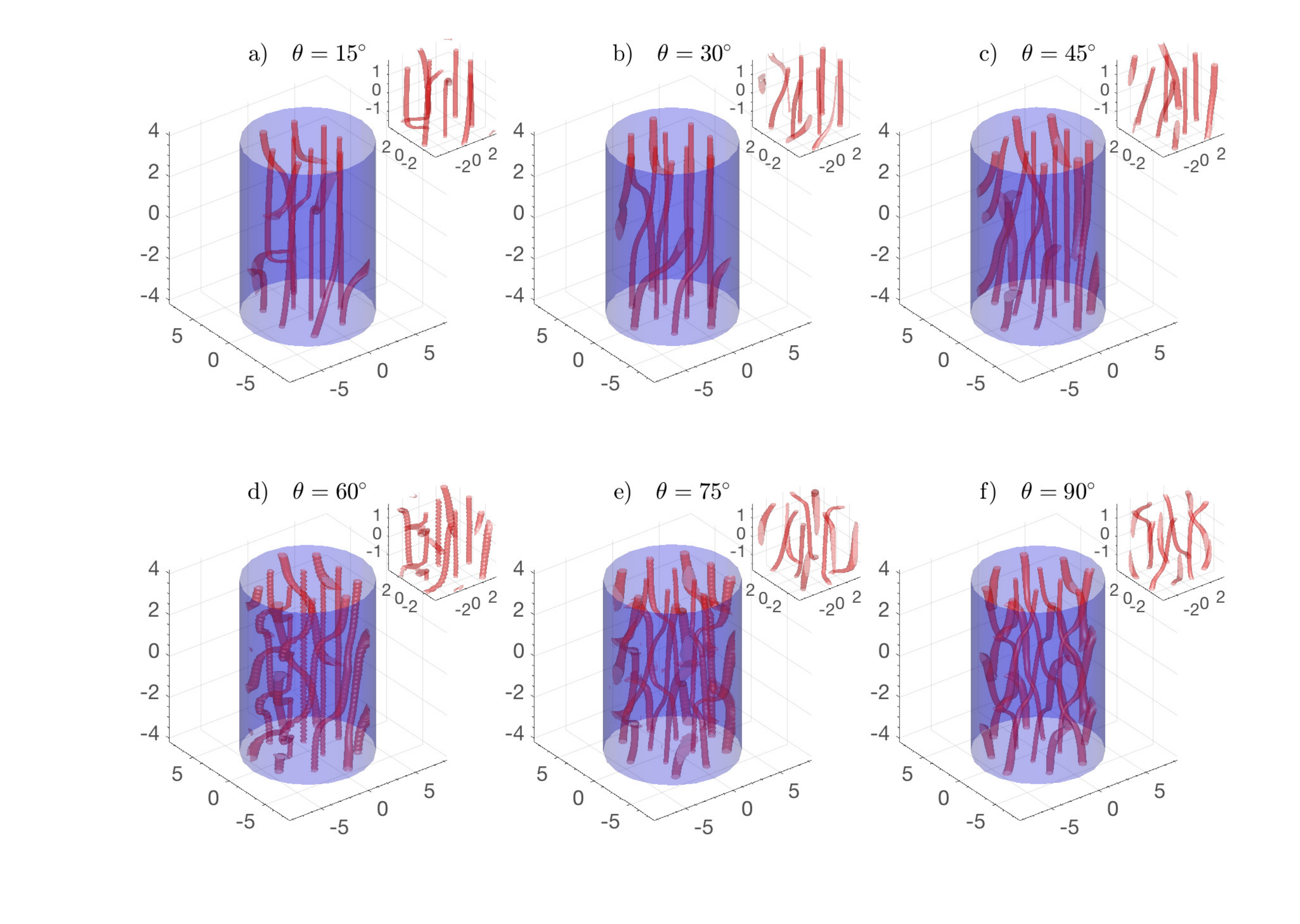}
    \caption{Three-dimensional ground-state structure of a neutron superfluid coupled to a proton superconductor containing an array of flux tubes with density coupling $\eta=-10$. The flux tube array is tilted at (a) $\theta=15^{\circ}$, (b) $\theta=45^{\circ}$, (c) $\theta=75^{\circ}$ and (d) $\theta=90^{\circ}$ with respect to $\boldsymbol{\Omega}$. Each panel displays the neutron density $|\psi|^2$ (in units of $n_n$), where the red shading signifies the vortex lines (drawn where $|\psi|^2$ drops below $10\%$ of its maximum in the core) and the blue shading marks the condensate's edge. Parameters: $\Omega=0.5$, $\tilde{N}_n=N_n/(n_n\xi^{3})=8\times10^3$, $d_{\Phi}=2$.}
    \label{fig:thetaplotwithinsets_small2}
\end{figure*}

\begin{figure}
\centering
	\includegraphics[width=0.85\columnwidth]{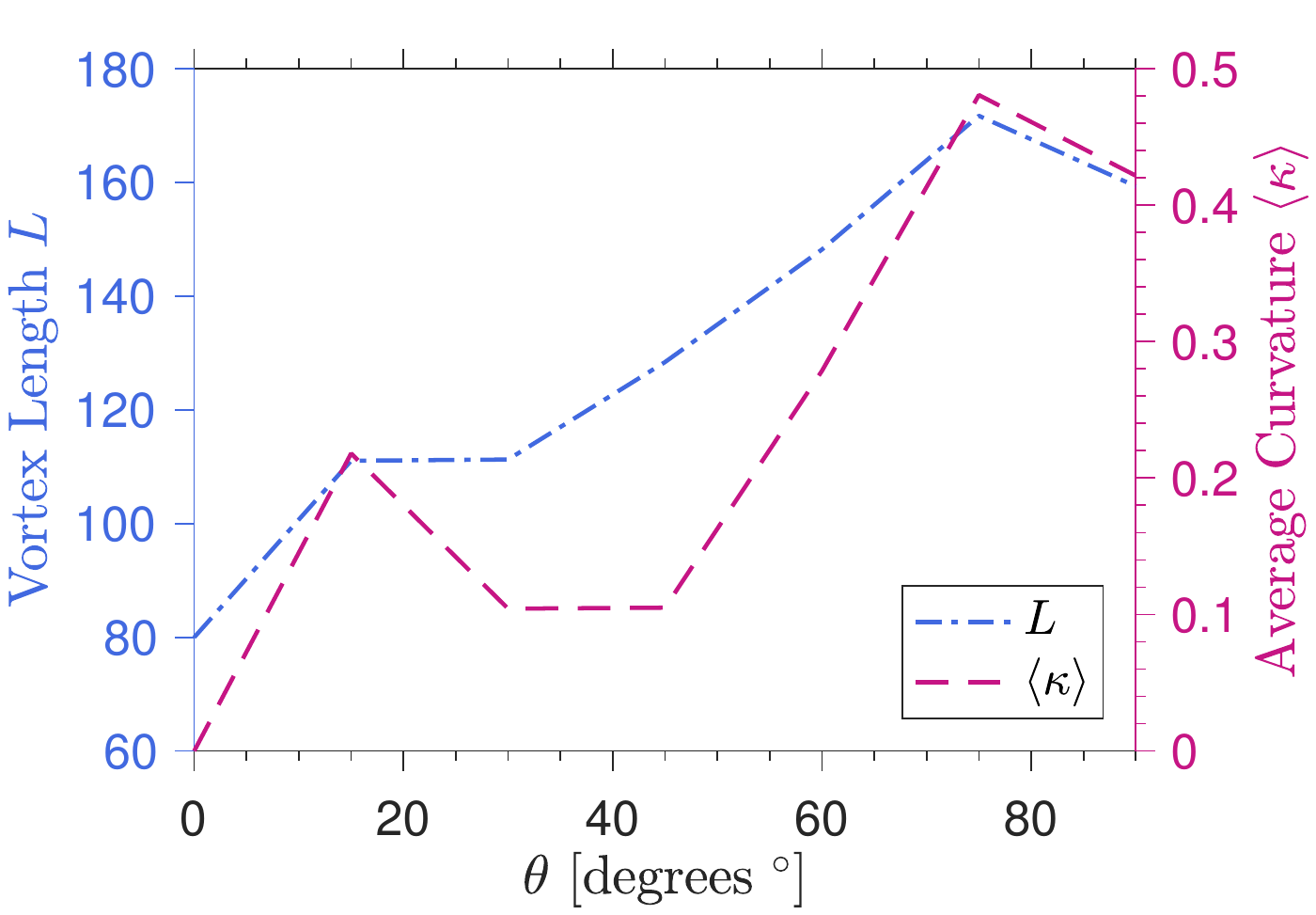}
    \caption{Vortex line length $L$ (in units of $\xi_n$) and average curvature $\langle\kappa\rangle$ (in units of $\xi_n^{-1}$) versus $\theta$ for the states plotted in Figure \ref{fig:thetaplotwithinsets_small2}.}
    \label{fig:linemeasures_plot}
\end{figure}

In this section, we suppose $\mathbf{B}$ makes an angle $\theta\neq0^{\circ}$ with respect to $\boldsymbol{\Omega}$. The special case $\theta=0^{\circ}$ is presented in Appendix \mbox{\ref{section:parallel}}. Figure \ref{fig:thetaplotwithinsets_small2} demonstrates that the vortex array is tangled, even when $\mathbf{B}$ is uniform. As $\theta$ increases, the number of close-to-straight vortices decreases; more vortices are tangled and bent microscopically  (see insets in Figure \ref{fig:thetaplotwithinsets_small2}). The competition between the two interactions produces zig-zag shapes: vortices align locally with the flux tubes but bend to align globally parallel to $\mathbf{\Omega}$ when they are more than $\xi_p$ away from the flux tube. The length-scales of the zigs and zags are therefore of order $d_{\Phi}$.  We observe that the sharpness of the zig-zag increases with $\theta$, so it takes more vortex length to span the condensate. Therefore, we expect to see $\langle \kappa \rangle$ and $L$ increase with increasing $\theta$. The states in Figure \ref{fig:thetaplotwithinsets_small2} have $\sim50$ flux tubes compared to $10$ vortices.

To quantify the behaviour in Figure \ref{fig:thetaplotwithinsets_small2}, we plot $L$ and $\langle \kappa \rangle$ versus $\theta$ in Figure \ref{fig:linemeasures_plot}. As expected, for  $\theta=0$, we have $\langle \kappa \rangle=0$ and $L=80.0$ (eight straight vortices each 10 units long). Both $L$ and $\langle \kappa \rangle$ increase as $\theta$ increases, as vortices try to align locally with $\mathbf{B}$ and globally with $\boldsymbol{\Omega}$. There are two deviations from this trend: "bumps" at $\theta=15^{\circ}$ and $\theta=75^{\circ}$. Between $\theta=15^{\circ}$ and $\theta=30^{\circ}$, L increases by $0.2\%$, while $\langle \kappa \rangle$ decreases by $52.2\%$. Examining the $\theta=15^{\circ}$ state visually in Figure \ref{fig:thetaplotwithinsets_small2}(a), we see a few sharply twisted vortices that boost $\langle \kappa \rangle$ in comparison to Figures \ref{fig:thetaplotwithinsets_small2}(b) and \ref{fig:thetaplotwithinsets_small2}(c). 

 Another interesting feature of Figure \ref{fig:linemeasures_plot} is the drop in both $L$ ($7.22\%$ decrease) and $\langle \kappa \rangle$ ($12.3\%$ decrease) between $\theta=75^{\circ}$ and $\theta=90^{\circ}$. We speculate that the drop occurs because the overlap length $2\xi_n/\mbox{sin} \ \theta$ at a vortex-flux-tube junction decreases, as $\theta$ approaches $90^{\circ}$ \citep{ChauDing,Link2012}. For this reason, even though $\theta=90^{\circ}$ may be the maximally frustrated state, it does not necessarily have the greatest $L$ or $\langle \kappa \rangle$. We need more resolution in $\theta$ and further studies with a larger system to fully investigate the above phenomena.

\begin{figure*}
	\includegraphics[width=\textwidth]{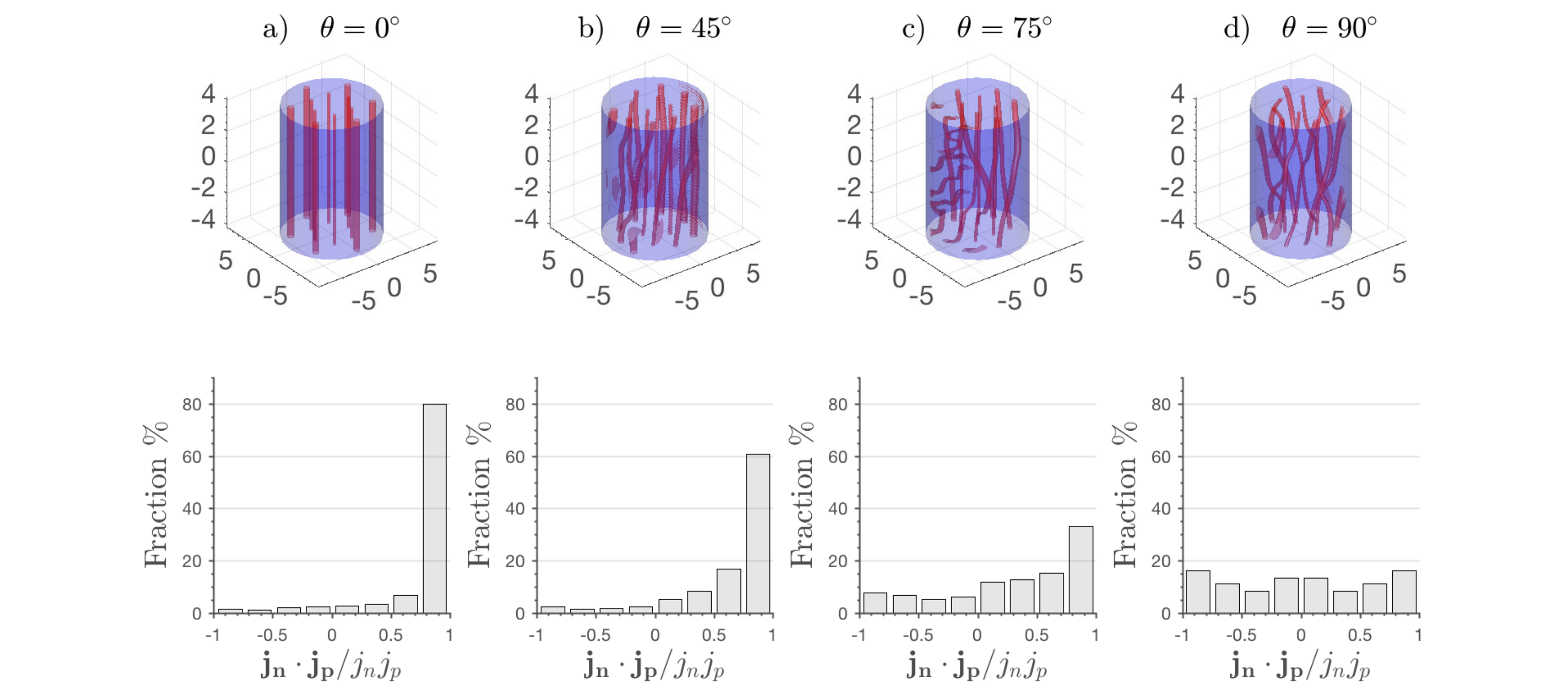}
    \caption{Three-dimensional ground-state structure of a neutron superfluid coupled to a proton superconductor containing an array of flux tubes with current-current coupling $\zeta=-1$. The flux tube array is tilted at (a) $\theta=0^{\circ}$, (b) $\theta=45^{\circ}$, (c) $\theta=75^{\circ}$ and (d) $\theta=90^{\circ}$ with respect to $\boldsymbol{\Omega}$. The top subpanels display the neutron density $|\psi|^2$ (in units of $n_n$), where the red shading signifies the vortex lines (drawn where $|\psi|^2$ drops below $10\%$ of its maximum in the core) and the blue shading marks the condensate's edge. The bottom subpanels display a histogram of $\mathbf{j}_n \cdot \mathbf{j}_p/j_nj_p$. Parameters: $\Omega=0.5$, $\tilde{N}_n=N_n/(n_n\xi^{3})=8\times10^3$, $d_{\Phi}=3$.}
    \label{fig:currentalignment3D}
\end{figure*}

\begin{figure}
	\includegraphics[width=0.95\columnwidth]{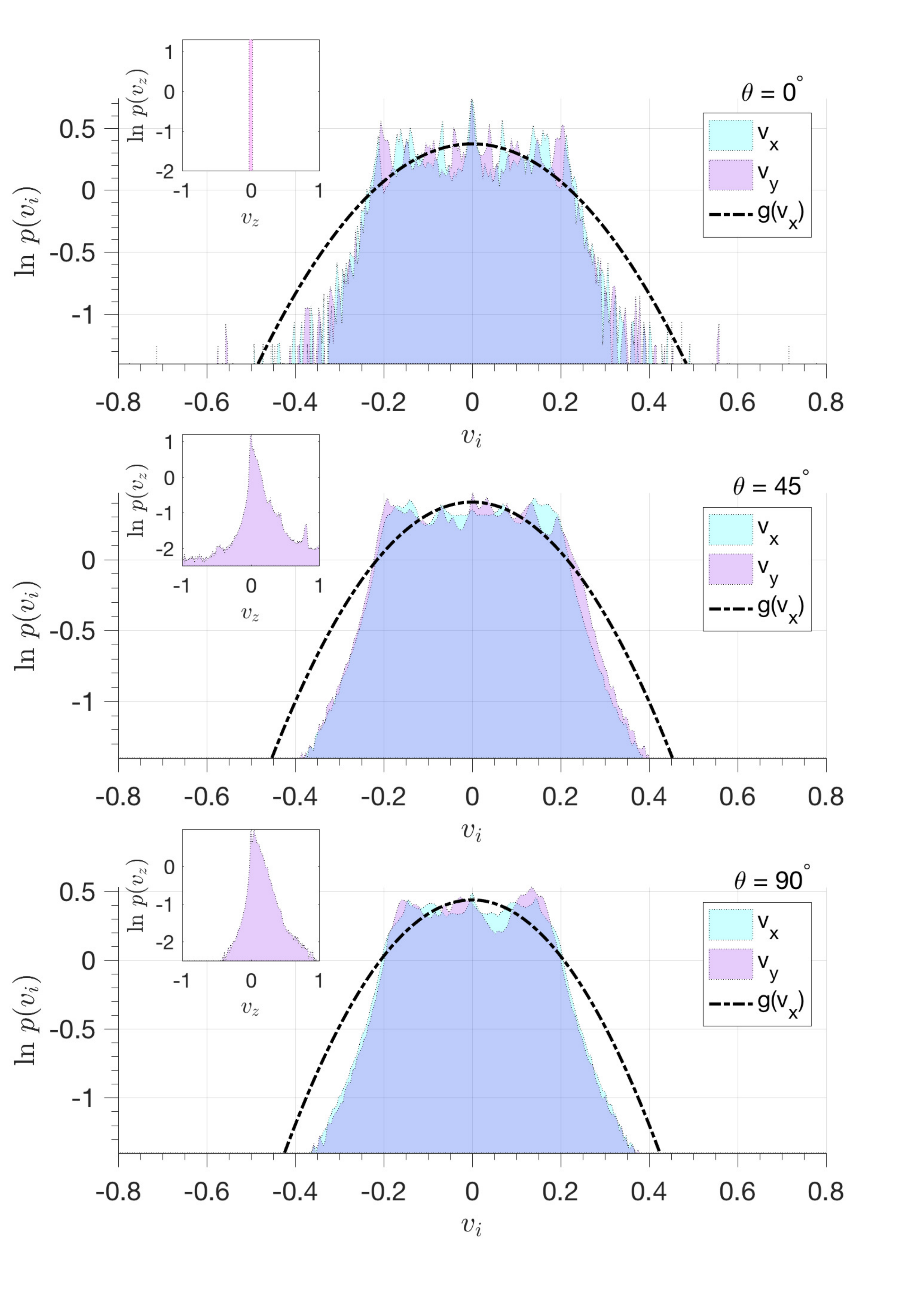}
    \caption{Logarithm of the probability density functions $p(v_i)$ of neutron velocity components $v_x$, $v_y$ and $v_z$, with $\mathbf{v}=\mathbf{j}_n/(m_n n_n)$ (in units of $c_s$), for three of the states pictured in Figure \ref{fig:currentalignment3D}. (\textit{Main panels}) $\ln p(v_x)$ and $\ln p(v_y)$ for $\theta=0^{\circ}$ (top), $\theta=45^{\circ}$ (middle) and $\theta=90^{\circ}$ (bottom). $\ln p(v_x)$ is shaded cyan and $\ln p(v_y)$ is shaded pink. The dotted curve is the logarithm of the gaussian probability density function $g(v_x)$, based on the mean $\mu_{v_x}$ and standard deviation $\sigma_{v_x}$ of the $x$-component of the velocity data.  (\textit{Inset}) $\ln p(v_z)$ for $\theta=0^{\circ}$ (top), $\theta=45^{\circ}$ (middle) and $\theta=90^{\circ}$ (bottom).}
    \label{fig:velocityhistogram}
\end{figure}

\subsection{Current alignment}
We are interested in the way $\theta$ controls the degree of frustration in the system. For example, for $\theta=90^{\circ}$, the vortices try to lie perpendicular to $\boldsymbol{\Omega}$, to promote current alignment, yet this would be disfavoured energetically, if there were no interaction present. Therefore, the currents are less likely to align as $\theta$ increases. In Figure \ref{fig:currentalignment3D} we plot four states featuring a current coupling of $\zeta=-1$ and $\theta=0^{\circ}$, $45^{\circ}$, $75^{\circ}$ and $90^{\circ}$. Figure \ref{fig:currentalignment3D} explicitly shows the frustration caused by the tilted flux tube array: as $\theta$ increases (i.e. moving to the right within the figure), the $\mathbf{j}_n\cdot\mathbf{j}_p/j_nj_p$ histogram becomes flatter, as the neutron currents find it harder to align with the proton currents. We observe qualitatively similar behaviour with a density coupling, i.e. the histogram flattens as $\theta$ increases for $\eta=-10$. The difference is that the average value of $\mathbf{j}_n\cdot\mathbf{j}_p/j_nj_p$ is larger for $\zeta=-1$, e.g. for $\theta=45^{\circ}$, we have $\langle\mathbf{j}_n\cdot\mathbf{j}_p\rangle=0.825j_nj_p$ for $\eta=-10$ and $\langle\mathbf{j}_n\cdot\mathbf{j}_p\rangle=0.910j_nj_p$ for $\zeta=-1$, where $\langle...\rangle$ denotes the average over all the grid points.

One might expect $E_{\mathrm{tot}}$ to rise due to an attractive coupling because, if $E_{\mathrm{self}}$ is nearly independent of $\theta$, $\mathbf{j}_n\cdot \mathbf{j}_p$ is lower and $E_{\mathrm{int}}$ is higher if the currents struggle to align. We verify that $E_{\mathrm{tot}}$ increases by $2.4\%$ from $\theta=0^{\circ}$ to $\theta=90^{\circ}$.

Finally, we plot probability density functions (PDFs) of the velocity at each grid point in Figure \ref{fig:velocityhistogram}, excluding locations outside the condensate's edge (drawn at the furthest distance from the axis, where $|\psi|^2$ drops below $10\%$ of its maximum). We compare the data with a gaussian PDF $g(v_x)$, based on the mean $\mu_{v_x}$ and standard deviation $\sigma_{v_x}$ of the $x$-component of the velocity data. We note two key features which are typical in quantum fluids: (i) "wiggles" in the PDF near $v_x=0$, probably caused by the anisotropy of the vortices \citep{White2010}; and (ii) the power-law tails associated with the singularity at the vortex cores in quantum fluids \citep{White2010}. The PDF converges to a Gaussian in the limit of a large number of vortices $N_v$ in the condensate ($N_v>10^6$)  \citep{White2010}, well beyond what our simulations can handle. The power-law tails for $\theta=0^{\circ}$, $\theta=45^{\circ}$ and $\theta=90^{\circ}$ have exponents $-7.11$, $-5.63$, $-5.27$  respectively. The standard deviation decreases with increasing $\theta$, e.g. $\sigma_{v_x}(\theta=0^{\circ})=0.170$, $\sigma_{v_x}(\theta=45^{\circ})=0.160$ and $\sigma_{v_x}(\theta=90^{\circ})=0.146$. In the $\theta=0^{\circ}$ case (top panel of Figure \ref{fig:velocityhistogram}), the flow is azimuthal, and we have $v_z=0$ (see inset of top panel). In the insets of the middle and bottom panels, for $\theta=45^{\circ}$ and $\theta=90^{\circ}$ respectively, the coupling to the tilted flux tubes introduces a non-zero $v_z$ component. This challenges the assumption that treating the mass currents of a neutron star as axisymmetric is accurate: flux tubes introduce superfluid velocity components along the rotation axis. If $v_z$ exceeds a threshold, the Donnelly-Glaberson instability can be excited \citep{Glaberson1974,donnelly1991,Peralta2005,Peralta2006}.

\section{Limitations of the Model}
\label{ref:limitations}
In this section, we discuss in further detail the idealizations inherent in the model noted in Sections \ref{section:GPmodel}--\ref{section:FTA}. The idealizations fall into three classes: (i) approximating the neutron superfluid in the outer core as a Bose-Einstein condensate coupled phenomenologically to a thermal reservoir and described by the stochastic GPE, discussed critically in Section \ref{section: GPEapprox}; (ii) assuming that the proton order parameter and magnetic flux tubes are static and prescribed, discussed critically in Section \ref{section:magneticansatz}; and (iii) working with a computationally tractable but astrophysically unrealistic dynamic range of parameter values, discussed critically in Section \ref{section:nsregime}. We emphasize again that caution must be exercised when interpreting the results of the simulations in an astrophysical context for the reasons detailed in this section and elsewhere.

\subsection{GPE approximation}
\label{section: GPEapprox}
GPE simulations have been employed previously with some success to describe glitch microphysics, e.g. vortex avalanches and the local, knock-on mechanisms that mediate them \citep{LilaMelatos2011, LilaMelatos2012,Douglass2015}. The latter studies generate avalanche statistics in line with observational data \citep{LilaMelatos2013}.

Nonetheless, it is important to point out that the GPE constitutes a simplification in the neutron star context. The Bogoliubov De Gennes (BdG) equations are appropriate for a weakly interacting Fermi gas, while the GPE describes a weakly interacting Bose gas. The matter in a neutron star fits neither of these descriptions completely: it is in the strongly interacting regime. Recently, the cross-over region applicable to strongly interacting Fermi gases has attracted attention in the literature \citep{Giorgini2008}. The cross-over region approaches the real situation in a neutron star, where the scattering length and average neutron-neutron spacing are of roughly the same order \citep{Chang2004}. In this regime, one finds a smooth transition from the BdG to the GPE \citep{Zwerger2016}. The latter results are most relevant in the unitarity limit, where the particles are both dilute and strongly interacting. Of course, the neutrons in the outer core of a neutron star are not dilute, but unitary Fermi gases are still believed to share some key properties with $^{1}S_{0}$ neutron superfluid matter \citep{Bulgac2017}. 

We do not solve the BdG equations in this paper due to computational limitations. The dimension of the BdG equations grows with the size of the system (i.e. a factor of $10^4-10^6$ increase in computational cost for our application) \citep{Han2010}. Solving the GPE is comparatively fast and captures certain properties of the neutron superfluid that we want to model (i.e. principally vortex dynamics).

Another simplification is the pairing state. Although we model the neutrons as an s-wave superfluid in this paper, it is sometimes argued that there is p-wave neutron pairing in the outer core, whereupon the order parameter becomes a 3 by 3 matrix in spin-angular-momentum space \citep{Yakovlev1999}.

\subsection{Magnetic ansatz}
\label{section:magneticansatz}
The magnetic vector potential is consistent with the proton order parameter (see Section \ref{section:protonSC}), but we are forced to neglect entrainment of the protons around the neutron vortices. We aim to address the proton response in future work but it is not easy; even with the static proton ansatz, meaningful three-dimensional runs require $\sim 1$ week of compute time. One important issue is that the potentially quite significant interaction of magnetic fields produced by neutron vortices and proton flux tubes is neglected. To address this problem in future, one must solve Ampere's law together with the proton and neutron equations of motion, making an already computing-intensive task even more demanding. It is likely that this will entail some trade-off with an even less realistic dynamic range than the present work accommodates.

An interesting question for the fully dynamical model is whether flux tubes can "lead" the interaction, i.e. whether the effect of the flux tubes on the neutron vortex structure is subordinate to the effect of the neutron vortices on the flux tube structure. The answer depends on the electron mean free path \citep{Harvey1986, Harrison1991, JahanMiri}. A static magnetic geometry is impossible in the presence of differential rotation if closed magnetic loops exist inside the star \citep{Easson1979, Melatos2012,GlampLasky}. Flux tubes can also creep outwards in a "leading" manner due to buoyancy forces \citep{MuslimovTsygan1985} or diffusion \citep{Guger2017}. 

\subsection{Parameter values}
\label{section:nsregime}
The realistic astrophysical parameter values pertinent to a neutron star are very different to those simulated in this paper. For example, the dimensionless size of the simulation box is $20$, corresponding to a dimensional value of $10^{-12}\ \mbox{m}$. A typical size of a pulsar is of order $10^4 \ \mbox{m}$. The number of vortices in the simulations is $N_v\approx 10\ \mbox{--} \ 80$ while in a pulsar the typical value of $N_v$ would be in the range $10^{16}\ \mbox{--} \ 10^{19}$.

Furthermore, the realistic parameter regime will remain inaccessible for many years due to computational limitations. In order to make a start, we follow a strategy used widely in other CPU-intensive problems: we order the relevant dimensionless parameters as in a neutron star and leave $\sim1 \ \mbox{dex}$ between them to give some separation in dynamic range for a manageable computational cost. In a wide variety of physical systems, the physical behaviour is qualitatively correct (a rescaled version of realistic behaviour), as long as dimensionless control variables have the correct relative ordering, e.g. the Reynolds and Prandtl numbers in stratified viscous flow \citep{clark_1973, friedlander_1976,duck_2001}.

The trapping potential $V$ plays multiple roles in our simulations. Its main purpose is practical: to stop the BEC from leaking out of the simulation volume and to keep it rotating. Its harmonic form is also broadly consistent with the potential resulting from hydrostatic balance in a neutron star ($V\propto r^{2}$ for a self-gravitating star with constant density). Some previous GPE simulations of the neutron star interior \citep{LilaMelatos2011, LilaMelatos2012} were conducted with hard-wall boundary conditions, while others \citep{Douglass2015} used a harmonic trap. The results are almost identical, successfully replicating vortex avalanche statistics in both cases. This is not surprising, as the exact shape of the trap does not feed into the microphysics of vortex unpinning. The same applies to the local interactions between proton flux tubes and neutron vortices studied in this paper.

\section{Conclusions}
This paper extends previous GPE simulations of the neutron star interior \citep{LilaMelatos2011, LilaMelatos2012,Douglass2015} in the following ways: (i) the system is three dimensional, allowing vortex bending and tangling to be investigated; (ii) the magnetic field is included by coupling the superfluid neutrons to the superconducting protons via density and current-current couplings in the presence of a prescribed, static flux tube array \citep{Clem}; and (iii) the angle between the rotation and magnetic field axes is arbitrary. It is found that the density and current-current interactions lead to broadly similar pinning behaviour, but current-current coupling rearranges the Abrikosov vortex lattice more extensively. Equilibrium configurations in three dimensions are "frustrated" and exhibit glassy relaxation characterised by "quakes" (sudden vortex rearrangement), as imaginary time elapses and the system moves between metastable configurations. The equilibrium configurations are tangled for $\theta\neq0^{\circ}$, with vortex length and mean curvature increasing with $\theta$. A component of superfluid velocity is induced along the rotation axis which can excite the Donnelly-Glaberson instability \citep{Glaberson1974} if it exceeds a critical threshold. These results call into question the assumption of rectilinear vortices in many neutron star models.

This problem is rich and many open questions remain. As an example, we pose the following astrophysically motivated question: do we still get a tangle (i.e. a frustrated equilibrium where the vortices try to align with both the magnetic and rotation axes in a competition between the local and global forces) in the limits $|\eta|\rightarrow\infty$ and $|\zeta|\rightarrow\infty$? Or do the vortices align with the flux tubes? The latter outcome seems impossible from an astrophysical standpoint, otherwise all neutron stars would have $\boldsymbol{\Omega}$ parallel to $\mathbf{B}$, which is not observed. The imaginary-time GPE simulations in this paper require $|\eta|$, $|\zeta|\lesssim2\times10^2$ for numerical stability, so we are unable to answer the question conclusively from first principles at the time of writing. Allowing for feedback onto the proton flux tubes in response to vortex forces may help resolve the apparent paradox. It is also relevant to investigate how the results vary with $n_{\Phi}$ in future work, particularly in the regime $n_{\Phi} \gg n_{v}$ pertinent to neutron stars. 

Another crucial question is to test what happens to the tangled equilibria in response to a spin-down torque, i.e. away from equilibrium, by solving the GPE in real time. Thus, we can self-consistently model the rotational dynamics and possibly simulate glitches. As the system spins down, vortices move radially outwards until they reach the edge of the condensate and leave the system. Do the flux tubes impede the outward vortex motion, as commonly assumed, or can the vortices slip past because they (and possibly the flux tubes) are tangled and even broken up into segments?

A first pass at answering the above question involves testing whether the tangled low-energy vortex states coupled to tilted flux tube arrays are less stable in response to a driving "push" than states with straighter vortices. We hypothesise that $\theta\neq0^{\circ}$ reduces the stability of the system. Results in Section \ref{section:FTA}, such as the increase in vortex length and mean curvature when vortices become tangled for $\theta\neq0^{\circ}$, are characteristic of instability in models of superfluid turbulence \citep{Barenghi2001,Peralta2006,AnderssonSidery2007}. It has been suggested that an instability occurs due to the stresses experienced by the vortices, when they are pushed against a rigid flux tube lattice \citep{Link2012}.

Secondly, by solving the Ginzburg-Landau equations of motion for the flux tubes simultaneously with the GPE for the vortices in future work, we can test how the flux tubes respond to the outward vortex motion driven by the star's spin down. Some possible behaviours are: (i) the flux tubes stay rigid and resist the vortices, until enough stress builds up for the vortices to break through; or (ii) the stress leads to vortex tangling and possibly segmentation, so that vortex loops slip past the flux tubes and vice versa. To explore the reaction of the flux tubes to the vortices, we need to allow to flux tubes to respond dynamically, which is beyond the scope of this paper. Unlike the static pinning sites associated with the nuclear lattice in the crust, the flux tubes are themselves embedded in a proton fluid which has its own complicated dynamics. Allowing the flux tubes to respond dynamically by solving the coupled Gross-Pitaevskii Ginzburg-Landau system would simulate the interaction of the two fluids more realistically and reveal the evolution of the magnetic field.

Three-dimensional GPE studies of neutron star spin down have the capacity to reveal new behaviours.  For instance, superfluid turbulence can lead to polarised vortex tangles \citep{Peralta2006,AnderssonSidery2007}. The unpinning probability can depend on whether the vortex unpins along its whole length or in small pieces \citep{LinkEpstein1991,LilaMelatos2011}. GPE simulations capture the effect of the fundamental anisotropy of the interaction between the two fluids; the orientation of the flux tubes with respect to the vortex motion is important \citep{Sauls1989,SideryAlpar}.

Macroscopic properties of the star such as the magnetic dipole moment $\mathbf{m}$ and angular velocity $\boldsymbol{\Omega}$ are intrinsically linked to the interaction of the vortex and flux tube arrays in the interior. Pinning to flux tubes is a possible mechanism for rotational glitches \citep{Sauls1989}. Dragging of the flux tubes by vortices as the star spins down can lead to magnetic field decay \citep{Srivasan1990}. Additionally, superfluid turbulence in the interior has astrophysical implications: it may explain the red timing noise observed in many neutron stars \citep{Link2012,MelatosLink2014} and emit stochastic gravitational radiation \citep{MelatosPeralta2010}. 

We emphasize again that the parameter regime and dynamic range pertinent to a neutron star are not accessible numerically at the time of writing, so it is challenging to link microscopic (vortex) scales to macroscopic (stellar) scales. To do this, the work needs to be extended to much larger systems, a formidable task.}


\section*{Acknowledgements}
The authors acknowledge helpful discussions with Tapio Simula in the early stages of this work and thank Michele Trenti for carefully reading the manuscript and catching a typographical error in equation (\ref{eq:curvature}). We thank the referee for their careful reading of our manuscript and constructive comments. Simulations were conducted using the MASSIVE cluster\footnote{\url{https://www.massive.org.au}} at Monash University with CPU time provided by a National Computational Infrastructure Merit Allocation Scheme grant. Support is also provided by the Australian Research Council through a Discovery Project grant and the Centre of Excellence for Gravitational Wave Discovery (OzGrav; CE170100004).



\bibliographystyle{mnras}
\bibliography{DrummondMelatos_revised2_clean} 

\appendix

\section{Modification of the kinetic energy term in the GPE due to entrainment}
\label{section:modification}
The dimensionless equation of motion describing the neutron order parameter $\psi$ is Equation (5) in \citet{AlparSaulsLanger}, viz.
\begin{equation}
\label{eq:gammaeq}
\mathrm{i}\frac{\partial\psi}{\partial t}=\left(-\tilde{\gamma}_{n}\nabla^{2}+V+|\psi|^{2}-\Omega\hat{L}_{z}\right)\psi+\mathcal{H}_{int}[\psi,\phi],
\end{equation}
where
\begin{equation}
\tilde{\gamma}_{n}=m_{n}/m_{n}^{*}
\end{equation}
modifies the effective mass in the kinetic energy term in \ref{eq:gammaeq} due to entrainment \citep{Andreev1975}. One typically finds $m_n\approx m_{n}^{*}$ in the outer core, according to various studies \citep{Sjoberg1976, entrainparam,Link2012}. Hence, we take  $\tilde{\gamma}_{n}\approx1$, obtaining
\begin{equation}
\label{eq:gammaeq2}
\mathrm{i}\frac{\partial\psi}{\partial t}=\left(-\nabla^{2}+V+|\psi|^{2}-\Omega\hat{L}_{z}\right)\psi+\mathcal{H}_{int}[\psi,\phi].
\end{equation}
As we neglect the proton response, only the kinetic energy term for the neutrons (which depends on $\tilde{\gamma}_{n}$) is explicitly present in Eq. (\ref{eq:gammaeq2}). The parameter $\tilde{\gamma}_{p}=m_{p}/m_{p}^{*}$ does enter the calculation through the static proton ansatz (see Appendix \ref{section:londoncoherence}).

\section{London penetration depth and proton coherence length}
\label{section:londoncoherence}
The proton superconductor has coherence length $\xi_p$ and London penetration depth $\lambda$. The proton coherence length is \citep{Mendell1998,Link2012}
\begin{align}
\xi_p&=16x_p^{1/3}\rho_{14}^{1/3}\frac{m_p}{m_p^*}\frac{1}{\Delta_p(\mbox{MeV})}\ \mbox{fm}\\
&=17\left(\frac{x_p}{0.05}\right)^{1/3}\left(\frac{\rho_{14}}{2.8}\right)^{1/3}\left(\frac{m_p/m_p^*}{2}\right)^{1/2}\\
&\ \ \ \times\left(\frac{\Delta_p}{60 \ \mbox{MeV}}\right)^{-1} \ \mbox{fm}  \nonumber
\end{align}
where $x_p=\rho_p/\rho_n$, $\rho_p$ is the proton mass density, $\rho_n$ is the neutron mass density, $\rho_{14}$ is the total mass density in units of $10^{14}\ \mbox{g cm}^{-3}$ and $\Delta_p(\mbox{MeV})$ is the proton pairing gap in $\mbox{MeV}$. The proton London penetration depth is \citep{AlparSaulsLanger,Link2012}
\begin{align}
\lambda&=30\left(\frac{m_p^*}{m_p x_p \rho_{14}}\right)^{1/2}\mbox{fm}\\
&=57 \left(\frac{x_p}{0.05}\right)^{-1/2}\left(\frac{\rho_{14}}{2.8}\right)^{-1/2}\left(\frac{m_p^*/m_p}{0.5}\right)^{1/2} \mbox{fm}.
\end{align}
Typical values for a neutron star are $\Delta_p=1\ \mbox{MeV}$  \citep{bandgap,Yakovlev1999}, $m_p/m_p^*\approx2$  \citep{Sjoberg1976,entrainparam,Link2012}, $x_p=0.05$ and $\rho_{14}=2.8$.

\section{Numerical value of $U_0$}
\label{section:numericalvalue}
The BEC healing length, which represents the typical distance over which spatial variations in $\psi$ occur, is given by \citep{Graber2016} 
\begin{equation}
\label{eq:BEC}
\xi_{\mbox{BEC}}=\frac{\hbar} {\sqrt{2\rho_n U_0}},
\end{equation}
where $\rho_n$ is the background mass density of the neutrons. In Bardeen-Cooper-Schrieffer (BCS) theory, the coherence length is comparable to the diameter of the vortex cores and is given by
\begin{equation}
\label{eq:BCS}
\xi_{\mbox{BCS}}=\frac{\hbar v_{Fn}} {\pi \Delta_n},
\end{equation}
where $v_F$ is the Fermi velocity and $\Delta_n$ is the energy gap for neutrons. The physical motivation for comparing (\ref{eq:BEC}) and (\ref{eq:BCS}) here is that we want the dimension of the BEC vortices to match those estimated in the neutron star literature using BCS theory. Writing $\xi_{\mbox{BEC}}\approx \xi_{\mbox{BCS}}$  semi-quantitatively, we find
\begin{align}
\label{eq:U0}
U_0&\approx\frac{\pi^2}{4}\left(\frac{\Delta_n^2}{E_{Fn}n_n}\right) \\
&=0.0051\left(\frac{\Delta_n}{0.1\ \mbox{MeV}}\right)^2 \left(\frac{n_n}{0.08\ \mbox{fm}^{-3}}\right)^{-1}\\
&\ \ \ \times \left(\frac{E_{Fn}}{60\ \mbox{MeV}}\right)^{-1} \ \mbox{MeV}\ \mbox{fm}^{3}.\nonumber
\end{align}
A typical value for the neutron energy gap is $\Delta_n\approx0.1\ \mbox{MeV}$  \citep{Beloin2016}. In (\ref{eq:U0}), $n_n$ is the number density of the neutron Cooper pairs and $E_{Fn}$ is the Fermi energy of the neutrons, with a typical value of $E_{Fn}=60\mbox{--}100\ \mbox{MeV}$  \citep{Shapiro,Yakovlev1999}.

\section{$\mathbf{B}$ parallel to $\boldsymbol{\Omega}$}
\label{section:parallel}
How do the vortex tangles computed in Sections \ref{subsection:VT} and \ref{subsection:glassy} depend on the angle between $\boldsymbol{\Omega}$ and $\mathbf{B}$? This question is answered in detail for $\theta\neq0^{\circ}$ in Section \ref{section:FTA}. In this appendix, we present the special case, where $\boldsymbol{\Omega}$ and $\mathbf{B}$ are parallel. This relatively simple example serves to demonstrate a few points. (i) Even when $\boldsymbol{\Omega}$ and $\mathbf{B}$ are parallel, there is a rich variety of complicated vortex configurations accessible in three dimensions. (ii) The square or triangular symmetry of the flux tube lattice is reflected in how the vortices bend. (iii) We cannot speak about a vortex being strictly pinned to or unpinned from a given flux tube, because it is partially pinned to multiple flux tubes.
\begin{figure}
\centering
	\includegraphics[width=0.9\columnwidth]{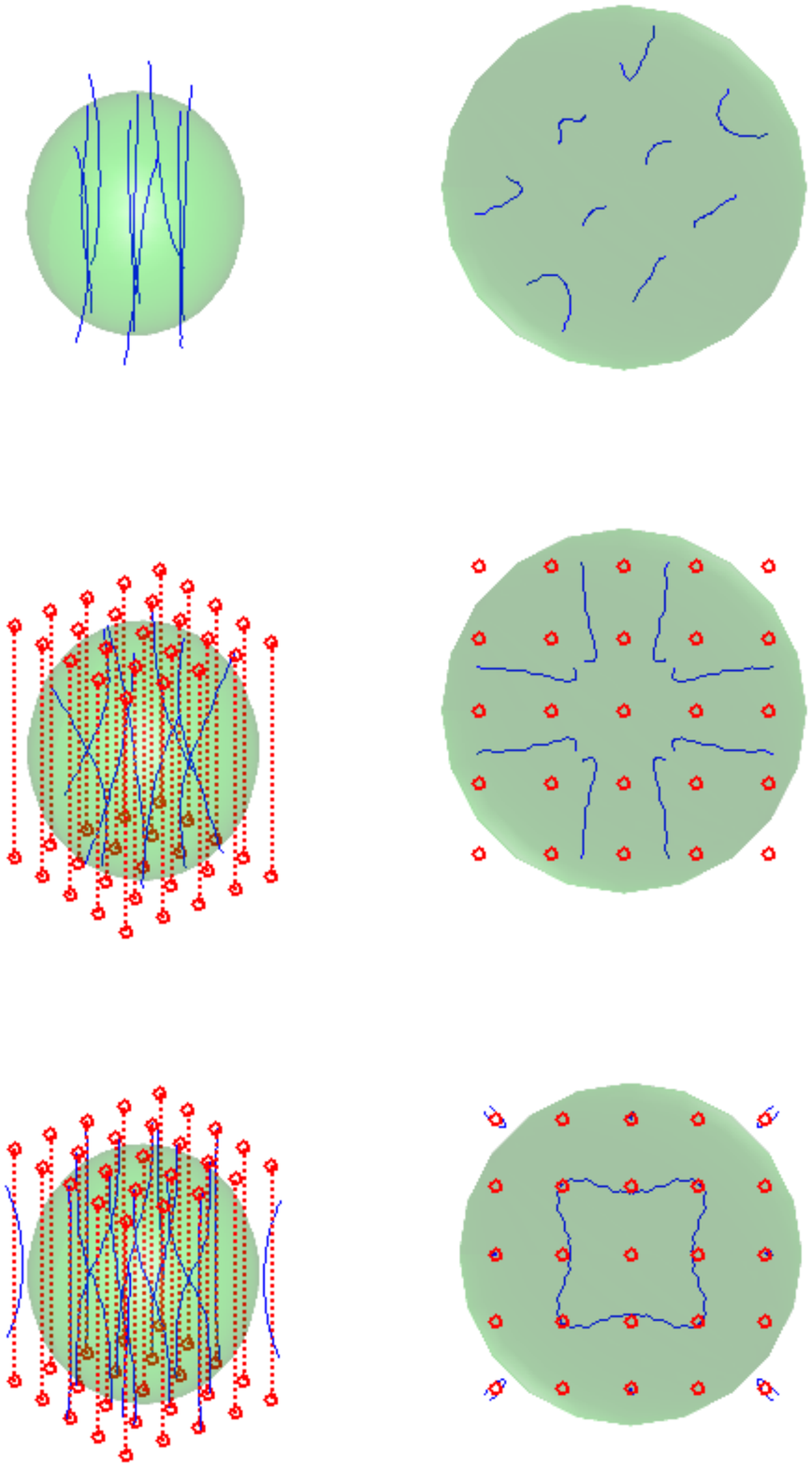}
\captionof{figure}{Ground-state vortex array configurations in three dimensions with $\theta=0^{\circ}$ as a function of density coupling strength given a square array of flux tubes. The green sphere is the Thomas-Fermi radius of the spherical condensate, the blue curves are the vortices (identified using the vortex finding algorithm described in Section \ref{subsection:VT}) and the red dotted curves are the flux tubes. (\textit{Top panel}) $\eta=0$. (\textit{Middle panel}) $\eta=10$. (\textit{Bottom panel}) $\eta=-10$. Parameters: $\tilde{N}_n=10^3$, $\Omega=0.7$, $d_{\Phi}=2$.}
    \label{fig:vortlines}
    \end{figure}
    
Figure \ref{fig:vortlines} displays ground-state vortex configurations with $\theta=0^{\circ}$ and reveals a wealth of new vortex configurations. In two dimensions, $\eta<0$ causes pinning, and $\eta>0$ causes vortices to sit between pinning sites. In three dimensions, by contrast, the optimal position for a vortex is between two flux tubes, so the vortex pins to the top half of one flux tube and the bottom half of the other as a compromise,  e.g. in the bottom image of Figure \ref{fig:vortlines}. Likewise, in the middle image of Figure \ref{fig:vortlines}, the repulsive interaction makes vortices curve around the flux tubes, so that they are as far away as possible from each flux tube and each other. The left side of Figure \ref{fig:vortlines} offers a side-on view, while the right side offers a bird's eye view. In the top right image (control state, i.e. zero coupling), the vortices form "squiggles" as they align along $\boldsymbol{\Omega}$, bending slightly because the spherical trap potential varies along their length. In the middle right panel ($\eta>0$), the vortices form a Maltese cross as they bend to avoid the flux tubes (marked by red circles). In the bottom right panel ($\eta<0$), the vortex lines join the dots to form a square shape as they pin partially to multiple flux tubes simultaneously. The square pattern copies the symmetry of the flux tube array.

Let us quantify the extent to which we get partial pinning of a vortex between two flux tubes. In order to do this, we first identify contiguous subsets of points along the vortex, as described in Section \ref{subsection:VT}. There are 25 flux tubes (plotted as red curves in Figure \ref{fig:vortlines}). We identify 16 contiguous vortex filaments (plotted as blue curves in Figure \ref{fig:vortlines}). For every point along each flux tube, we scan over every point along each vortex. If the vortex point lies within $1.4\xi_p$ of the flux tube point, we classify it as pinned to that flux tube. Thus, we confirm numerically what we see in Figure \ref{fig:vortlines}: nine  flux tubes have no vortices pinned to them, eight of the flux tubes have a single vortex pinned to them, and eight have segments of two different vortices pinned to them. 

Among the eight flux tubes with one vortex pinned to them, four have $\approx130$ vortex points pinned and four have $\approx60$ vortex points pinned. Among the eight flux tubes with two vortices pinned, all have $\approx80$ vortex points pinned, with about half contributed by each pinned vortex (maximum ratio $=56 \%:44 \%$). This example illustrates partial pinning of a vortex between flux tubes, triggered by variation in density along the direction of $\boldsymbol{\Omega}$ in a neutron star. There are two possible ways of viewing the role of the spherical trap, relevant to different length scales: (i) the spherical trap simulates microscopic variation in forces along $\boldsymbol{\Omega}$ due to, for example, random fluctuations; or (ii) Figure \ref{fig:vortlines} presents a toy demonstration of macroscopic bending on stellar scales of the lines in the presence of a flux tube array.

\bsp	
\label{lastpage}
\end{document}